\documentclass[a4paper,twoside]{article}
\usepackage[latin1]{inputenc} 
\usepackage[T1]{fontenc} 
\usepackage{RR}
\usepackage{hyperref}

\usepackage{paralist}
\usepackage{graphicx}
\usepackage{subfigure}
\usepackage{listings}
\usepackage{amsmath}
\usepackage{wasysym}
\usepackage{amssymb}
\usepackage{amsfonts}
\usepackage{moreverb}
\usepackage{url}
\usepackage{cite}
\usepackage{multirow}
\usepackage{hyperref}
\usepackage[all,2cell,ps]{xy}

\newtheorem{definition}{Definition}[section]
\newtheorem{theorem}{Theorem}

\usepackage[T1]{fontenc}
\usepackage{alltt}
\usepackage{verbatim}
\usepackage{moreverb}
\usepackage{amssymb}
\usepackage{listings}
\usepackage{float}
\usepackage[linesnumbered,vlined]{algorithm2e}
\usepackage{algorithmic}

\usepackage{graphicx}
\usepackage{wrapfig}

\newenvironment{algo}{%
\algorithm
}{%
\endalgorithm
}

\RRNo{7834}
\RRdate{December 2011}
\RRauthor{
Houari Mahfoud \thanks{Univ. Nancy 2, UMR 7503 \&  INRIA Nancy Grand Est 
(\texttt{Houari.Mahfoud@inria.fr}).} and
Abdessamad Imine\thanks{Univ. Nancy 2, UMR 7503 \&  INRIA Nancy Grand Est
(\texttt{Abdessamad.Imine@inria.fr}).}
}
\authorhead{H. Mahfoud and A. Imine}
\RRtitle{Interrogation S\'ecuris\'ee des Vues XML R\'ecursives:\\Une Technique bas\'ee sur le Standard XPath}
\RRetitle{Secure Querying of Recursive XML Views:\\A Standard XPath-based Technique} 
\titlehead{Secure Querying of Recursive XML Views:\\A Standard XPath-based Technique}
\RRresume{
La plupart des travaux existant autour du contr\^ole d'acc\`es des documents XML se basent sur la d\'efinition d'une vue 
pour chaque utilisateur qui repr\'esente les parties des donn\'ees dont il est autoris\'e \`a lire et/ou modifier. 
Cette vue est le r\'esultat de 
l'annotation de la grammaire associ\'ee au document XML (par exemple une DTD) par diff\'erentes conditions 
d'acc\`es exprim\'ees sous forme d'expressions XPath. Pour emp\^echer l'acc\`es \`a des donn\'ees confidentielles -- cach\'ees par la vue --, chaque requ\^ete  pos\'ee par l'utilisateur sur 
la vue doit \^etre r\'e\'ecrite pour qu'elle soit \'evalu\'ee en toute s\'ecurit\'e sur le document original. Cette r\'e\'ecriture permet d'\'eviter
le co\^ut de la mat\'erialisation et de la maintenance de la vue.
Cependant, la r\'e\'ecriture des requ\^etes XPath dans le cas des vues XML 
r\'ecursives reste un probleme ouvert. Pour pallier \`a ce probl\`eme,
certains travaux ont propos\'e de travailler avec un langage de requ\^etes non-standard, appel\'e ``Regular XPath''. 
N\'eanmoins, le langage ``Regular XPath'' reste au stade th\'eorique car 
aucun outil d'\'evaluation n'est disponible en pratique. Une impl\'ementation de ce langage est bas\'ee sur les automates, ce qui peut engendrer une complexit\'e de r\'e\'ecriture exponentielle.\\
Dans ce papier, nous montrons que la r\'e\'ecriture des requ\^etes XPath dans le cas des vues XML r\'ecursives est possible sans passer par des transformations vers 
d'autres langages (tel que ``Regular XPath''), et peut \^etre faite en temps lin\'eaire. 
Nous \'etudions l'extension du fragment XPath, appel\'e en anglais \emph{downward class} (compos\'e seulement par les axes \emph{child} et \emph{descendant}), 
par certains axes et op\'erateurs XPath. 
En nous basant sur cette extension, nous proposons un mod\`ele g\'en\'eral pour r\'e\'ecrire des requ\`etes XPath pour des vues XML arbitraires, r\'ecursives ou non. 
Une phase d'exp\'erimentation montre bien l'efficacit\'e ainsi que le passage \`a l'\'echelle de notre algorithme de r\'e\'ecriture.}

\RRabstract{
Most state-of-the art approaches for securing XML documents allow users to access data only through authorized
views defined by annotating an XML grammar (e.g. DTD) with a collection of XPath expressions. To prevent improper
disclosure of confidential information, user queries posed on these views need to
be \emph{rewritten} into equivalent queries
on the underlying documents. This rewriting enables us to avoid the overhead of view materialization and maintenance.
A major concern here is that query rewriting for recursive XML views is still an
\emph{open} problem. To overcome this
problem, some works have been proposed to translate XPath queries
into non-standard ones, called Regular XPath
queries. However, query rewriting under Regular XPath can be of
exponential size as it relies on automaton model.
Most importantly, Regular XPath remains a theoretical achievement. Indeed, it is
not commonly used in practice as translation and
evaluation tools are not available.
In this paper, we show that query rewriting is always possible for recursive XML views using only the expressive power of the standard XPath.
We investigate the extension of the downward class of XPath, composed only by \emph{child} and \emph{descendant} axes, with some axes and operators and we propose a general approach to rewrite queries under recursive XML views. 
Unlike Regular XPath-based works, we provide a rewriting algorithm which processes the query only over
the annotated DTD grammar and which can run in linear time in the size of the query. An experimental evaluation demonstrates that our
algorithm is efficient and scales well.
}
\RRmotcle{R\'e\'ecriture des requ\^etes, Contr\^ole d'acc\`es pour documents XML, Vues XML, XPath.}
\RRkeyword{Queries Rewriting, XML Access control, XML views, XPath.}
\RRprojets{CASSIS}
 \RCNancy 

\begin{document}
\makeRR   
\tableofcontents
\newpage

\section{Introduction}
XML has become the standard of representation and exchange of data across the web. With this
emergence, a challenge is raised with regards to the security of XML documents whose content is available to one or more users based on their access privileges. First access control models for securing XML have been proposed in~\cite{ref2,ref1,ref12}.
However, these models suffer from various limitations. 
They can cause leakage of sensitive information \cite{ref2}, focus on the annotation of the entire XML data to deal with the static analysis 
limitations \cite{ref1}, or based on costly schemes for rewriting user queries \cite{ref12}. 

To avoid these problems, the notion of XML security views was studied by Fan et al. \cite{ref4}. We briefly review the main principle of the XML 
security view-based approaches. 
We start by the fragment of XPath, called \emph{downward} class, where the axes are limited to \emph{child} and 
\emph{descendant} axes. We use this fragment since it is commonly used in practice.
A conform  XML document \emph{T} w.r.t  a DTD \emph{D} (i.e. \emph{T} is an instance of \emph{D}), can be queried simultaneously by different users.
For each class of users a security view is defined by annotating \emph{D} with some access conditions 
to specify the (in)accessible element types of the DTD. The annotated version of \emph{D} is later sanitized by removing the inaccessible element types which 
results in a DTD view $D_v$. Then the security view is defined as \emph{V}=$(D_v,\sigma)$ where $D_v$ 
is given to the users, which describes accessible data they are able to see, and $\sigma$ 
is a function used to extract for each XML document \emph{T} conforms to \emph{D}, its 
view $T_v$ representing only authorized data of the users. Each query over the view $T_v$ is translated into an equivalent one in order to be evaluated over the original data \emph{T}.\newline

\noindent \textbf{Problem Statement.}
The problem of XPath queries rewriting studied in this paper is defined as follows:
\begin{center} 
\fbox{\begin{minipage}{4.5in}
\noindent Given a DTD $D$, an XML security view $V$=$(D_v,\sigma)$, and an XPath query $Q$ over $D_v$. The rewriting problem consists in defining 
a rewriting function $\mathcal{R}$ that computes another XPath query $\mathcal{R}(Q)$ over the original document $D$ such that: 
for any instance $T$ of $D$ and its view $T_v$ computed w.r.t $V$, the evaluation of $Q$ on $T_v$ yields the same result as the evaluation 
of $\mathcal{R}(Q)$ on T.
\end{minipage}} 
\end{center}

\noindent Most of the security view-based approaches of XPath queries rewriting deal only with non-recursive DTDs \cite{ref4,ref6,ref7}. 
A DTD is recursive iff at least one of its elements is defined (directly or indirectly) in terms of itself. Note that 
recursive DTDs often arise when specifying medicals data and the problem of query rewriting is more intriguing in this case. 
A security view is recursive, if its view $D_v$ is recursive.

For each pair of element types \emph{A} and \emph{B} in the DTD, the $\sigma$ function is an XPath expression denoting the set 
of paths to reach an element \emph{B} from an element \emph{A} in the DTD view  (where some element types are hidden between \emph{A} and \emph{B} 
and generated by $\sigma$). 
However, in the case of recursive DTDs, the $\sigma$ function is not computable since 
there may be an infinite set of paths from \emph{A} to \emph{B} and the notion of security view (as defined above) cannot be used in 
queries rewriting.
That is why some authors \cite{ref5,ref8} resort to the use of Regular XPath to avoid this problem. However, Regular XPath
remains of theoretical use since no evaluation tools have been provided for practical use of this langage. 

To the best of our knowledge, no practical approach exists for 
answering queries under recursive XML security views. Accordingly, the XPath query rewriting remains an open issue.\newline

\noindent \textbf{Contribution.} 
Our main contribution is making possible the query rewriting for recursive XML security views using only 
the expressive power of the standard XPath.
We show that  extending the downward class of XPath queries with 
some axes and operators is sufficient to deal with the query rewriting  under recursion (without the need of the \emph{Kleene star} or the translation from XPath to Regular XPath). 

Intuitively, for a query \emph{Q} based on the downward class of XPath, our rewriting solution consists in computing another query 
$Q'=\mathcal{R}(Q)$ using an extended fragment of XPath in such a way for any instance \emph{T} of \emph{D} and its view $T_v$
(computed w.r.t $D_v$) the evaluation of \emph{Q} on $T_v$ gives the same result than evaluating $Q'$ on \emph{T}.

We provide a linear rewriting algorithm for arbitrary views (recursive or not) which,
unlike Regular XPath-based works (relying on Mixed Finite Automata\cite{ref5}),
 consists only in processing the query over the annotated DTD to produce the equivalent query on any valid instance
of the original DTD. We validate our solution with a performance evaluation which shows that our rewriting algorithm
is efficient and scales well.
Lastly, we show how our proposed solution can be extended to deal with a large fragment of XPath (including upward-axes)
and to go beyond some limitations of existing access control specification languages.\newline

\noindent \textbf{Related Work.}
We briefly discuss two approaches of access control policy enforcement for XML documents with or  without XML grammar.

In \cite{ref2} authors propose a formal model to specify access control for XML documents independently of the DTD. The policy rules definition is based on XPath 
and each query is rewritten by adding a predicate access (which represents all accessible data) to that query. 
We use the same predicate access principle in our rewriting approach.
However, inference of sensitive information can be detected since only the last subquery is controlled among all subqueries 
parsed by the query. To overcome this problem, we improve the method given in\cite{ref2}  by attaching a predicate access to each entity (element/attribute) parsed by the query.

Vercammen \cite{ref12} proposes a new method based on the intersection and union of XPath queries to avoid the 
problem of information leakage. 
The policy rules are translated to a single query which stands for all accessible data; this query is incorporated by intersection with each query requested over the user XML document view. However, this approach yields the same performance than the materialization of this view.

Other access control approaches are based on the notion of security views and the query rewriting principle. 
Fan et al. \cite{ref4} propose the notion of security view by the 
annotation of a regular non-recursive DTD. The use of only downward class of XPath queries 
allowed them to achieve more precise query rewriting, i.e. computing all possible paths connecting each two adjacent elements in the query, which provides practically performance gains for the query evaluation. 
A view derivation algorithm is proposed to compute the DTD view, w.r.t. the access conditions, and an optimization 
step is also done over the rewritten query. However, to keep the DTD view regular, an inaccessible element may be replaced with 
anonymous element \emph{dummy} which can be source of 
security breaches. In \cite{ref6,ref7}, authors refine the Fan's model  by eliminating \emph{dummies}, extending the class of XPath queries with \emph{upward}-axes and 
with a novel notion of security views. Different types of policies are also discussed. These works can deal only with non-recursive DTDs.
They are inapplicable to recursive DTDs because the description of paths connecting two element types may be infinite.

Unlike the XPath query rewriting over non-recursive DTDs, the problem posed by the recursion has not received a more attention. 
Authors of \cite{ref5} extend the principle proposed in \cite{ref4} with a translation of XPath queries to Regular XPath and propose a first algorithm for 
evaluating Regular XPath over XML data. In \cite{ref8}, a more generalized rewriting approach has been studied by dealing with restrictions on the class of queries 
and DTD types. The defined accessibility function is based on the \emph{Kleene star}. It should be noted that  the \emph{Kleene star} cannot be expressed in the standard XPath.

Although the query formulation and rewriting on Regular XPath is more expressive than the standard XPath, we cannot find any practical system for both proposed approaches\footnote{According to \cite{ref13} the SMOQE system proposed in \cite{ref15} has been removed because of conduction of future researches.}. 
Consequently, the need of a rewriting system of XPath queries over recursive XML security views remains an open issue.\newline

\noindent \textbf{Plan of the paper.} 
The rest of the paper is organized as follows. Section \ref{Sect2} presents formally the query rewriting problem for
recursive views, and sketches our solution to deal with this problem. In Section \ref{Sect3}, we give the ingredients of our
access control specification. Our rewriting approach is detailed in Section \ref{Sect4}. Section \ref{extensions} presents how our approach can
be extended to consider a large fragment of XPath and used to overcome some limitations of existing access control approaches. 
An implementation issue is presented in Section \ref{Sect6}. Finally, we conclude this paper in Section \ref{Sect7}.

\section{Formal Problem Statement}\label{Sect2}
In this section we present the query rewriting problem for recursive views, and sketch our solution to deal with this problem.

\subsection{Preliminaries}
We briefly review some notions of Document Type Definitions (DTDs) and the class of XPath Queries most used in practice.\newline

\noindent\textbf{DTDs.} Without loss of generality, we represent a DTD by a triple (\emph{Ele}, \emph{P}, \emph{root}), where \emph{Ele} is a finite set 
of \emph{element types}, \emph{root} is a 
distinguished type in \emph{Ele} (called the \emph{root type}) and \emph{P} is a function defining element types such that for any \emph{A} in 
\emph{Ele}, \emph{P(A)} is a regular expression $\alpha$ defined with:

\begin{center}
$\alpha$ := $str$ | $\epsilon$ | $B$ | $\alpha$","$\alpha$ | $\alpha$"|"$\alpha$ | $\alpha$*
\end{center}

\noindent where \emph{str} denotes the text type \texttt{PCDATA}, $\epsilon$ is the empty word, \emph{B} is an element type in \emph{Ele}, and finally 
$\alpha$","$\alpha$, $\alpha$"|"$\alpha$, and $\alpha$* denote concatenation, disjunction, and the Kleene closure respectively. 
We refer to $A \rightarrow P(A)$ as the \emph{production} of \emph{A}. For each element type \emph{B} occurring in \emph{P(A)}, we refer to 
\emph{B} as a \emph{subelement type} (or \emph{child type}) of \emph{A} and to \emph{A} as a \emph{superelement type} 
(or \emph{parent type}) of \emph{B}. If an element type \emph{A} is defined in term of an element type \emph{B} directly (\emph{B} is subelement type of \emph{A}) or indirectly, 
then \emph{A} is an \emph{ancestor type} of \emph{B} and \emph{B} is a \emph{descendant type} of \emph{A}. The DTD is said \emph{recursive} if some element type \emph{A} is 
defined in terms of itself directly or indirectly.

We use the graph representation to depict our DTDs where dashed edges represent disjunction.\newline

\noindent \textbf{XML Documents.} We model an XML document with an unranked ordered finite node-labeled tree, also called \emph{XML Tree}. Let $\Sigma$ be a 
finite set of node labels, an XML tree \emph{T} over 
$\Sigma$ is a structure defined as\cite{ComplexityXPath20}: $T=(N,R_{\downarrow},R_{\rightarrow},L)$ where \emph{N} is the set of nodes, 
$R_{\downarrow}$ $\subseteq$ $N\times N$ is the parent-child relation, 
$R_{\rightarrow}$ $\subseteq$ $N\times N$ is a successor relation on (ordered) siblings, and $L: N\rightarrow \Sigma$ assigns a label to each node.

An XML document \emph{T} conforms to a DTD \emph{D} if the following conditions hold: (i) the root of \emph{T} is the unique node labeled with \emph{root}; 
(ii) each node in \emph{T} is labeled either with element type \emph{A}, called \emph{A element}, or with \emph{str}, called 
\emph{text node}; (iii) for each node \emph{n} of type \emph{A} and with \emph{k} ordered children $n_1,...,n_k$, the word $L(n_1),...,L(n_k)$ belongs to the regular 
language defined by \emph{P(A)}; (iv) each text node carries 
a string value (\texttt{PCDATA}) and is the leaf of the tree. We call \emph{T} an instance of \emph{D} if \emph{T} conforms to \emph{D}. 
In the DTD instances depicted in our figures, we use $X^{i}$ to distinguish between elements of the same type \emph{X}.\newline

\label{XPathQueries}
\noindent\textbf{XPath Queries.} We introduce the \emph{downward} class of XPath queries referred to as $\mathcal{X}$ and defined as follows:
\begin{alltt}
 \texttt{path}    :=  \texttt{axis}::\texttt{label} | \texttt{path}'['\texttt{qual}']'
            | \texttt{path}'/'\texttt{path} | \texttt{path}'\(\cup\)'\texttt{path}
 \texttt{qual}    :=  \texttt{path} | \texttt{path} = \(c\)
            | \texttt{qual} \(and\) \texttt{qual} | \texttt{qual} \(or\) \texttt{qual} 
            | \(not\) \texttt{qual} | '('\texttt{qual}')'
 \texttt{axis}    :=  \(\downarrow\) | \(\downarrow\sp{+}\)
\end{alltt}

\noindent where \emph{label} refers to element type in \emph{Ele} or \texttt{*} (that matches all labels), $\cup$ stands for union, 
\emph{c} denotes \emph{text constant}, \emph{axis} is the XPath axis relation, and $\downarrow, \downarrow^{+}$ denote \emph{child} and \emph{descendant} axis respectively, 
and finally $qual$ is called 
an XPath \emph{qualifier} (\emph{predicate} or \emph{filter}) which can be a text content comparison, an XPath query, or a boolean 
expression (using boolean operators such as: \emph{and}, \emph{or}, \emph{not}). 

Let \emph{n} be a node in an XML tree \emph{T}. The evaluation of an XPath query \emph{p} at node \emph{n}, called \emph{context node n}, results in a set of 
nodes which are reachable from \emph{n} with \emph{p}, denoted by \emph{n}\textlbrackdbl$p$\textrbrackdbl. A qualifier \emph{q} is said valid at 
node context \emph{n}, denoted by $n\vDash q$, iff one of the following conditions holds: (i) \emph{q} is given by \emph{p=c} and there is at least 
one element reachable from \emph{n} with \emph{p} which has \emph{c} as text content; (ii) \emph{q} is an XPath query and \emph{n}\textlbrackdbl$q$\textrbrackdbl$\;$ is nonempty; 
(iii) \emph{q} is a boolean expression (e.g. \emph{not(p)}) and it is evaluated to true at \emph{n}.

\subsection{XML Access Control Model}
We define below some concepts of security specifications and XML security views as initially presented 
in~\cite{ref4,ref6}.\newline

\noindent \textbf{Security Specifications.} Given an XML document \emph{T} conforms to a DTD \emph{D}. For each class of users some access privileges may be defined to restrict access to sensitive information on \emph{T}. 
Thus, an access-control specification language is defined to specify what elements in \emph{T} the users are granted, denied, or conditionally granted access to. An \emph{access specification} in this 
language is defined as follows:

\begin{definition}\label{accessSpecification}
An access specification \emph{S} is a pair \emph{(D,\emph{\texttt{ann}})} consisting of a DTD \emph{D} and a partial mapping \emph{\texttt{ann}} such that, for each production $A \rightarrow P(A)$ and 
each element type \emph{B} in \emph{P(A)}, \emph{\texttt{ann($A$,$B$)}}, if explicitly defined, is an annotation of the form: 
\begin{center}
 \emph{\texttt{ann($A$,$B$) :=  $Y$ | $N$ | [$Q$]}}
\end{center}
where \emph{\texttt{[$Q$]}} is a qualifier in our XPath fragment $\mathcal{X}$. A special case is the root of \emph{D} for which we define \emph{\texttt{ann}}$(root)$=$Y$ by default.\hfill\(\square\)
\end{definition}

\noindent The specification values \emph{Y}, \emph{N}, and [$Q$] indicate that the \emph{B} children of \emph{A} elements in an XML document conforms to the DTD \emph{D} 
are \emph{accessible}, \emph{inaccessible}, or \emph{conditionally accessible} respectively. If \texttt{ann}(\emph{A,B}) is not explicitly defined, then \emph{B} inherits 
the accessibility of \emph{A}. On the other hand, if \texttt{ann}(\emph{A,B}) is explicitly defined then \emph{B} may 
\emph{override} the accessibility inherited from \emph{A}. A text node is accessible only if its parent element is accessible.
For an element node \emph{n} of type \emph{B} with parent node of type \emph{A}, we say that \emph{n} is concerned by an annotation if \texttt{ann}(\emph{A,B})=\emph{value} exists, moreover, 
this annotation is valid at \emph{n} if \emph{value=Y}, or \emph{value}=[\emph{Q}] and $n \vDash Q$.\newline

\noindent \textbf{Security Views.}
To enforce  an access specification, a \emph{security view} is defined to compute for each document \emph{T} conforms to a DTD \emph{D}: 
(i) an instance view $T_v$ containing only accessible data; and, (ii) a DTD view $D_v$ which describes 
schema of all accessible data. Both documents $T_v$ and $D_v$ are seen only by authorized users.

More formally, let \emph{S}=$(D,\texttt{ann})$ be an access specification. A security view for \emph{S} is defined as a pair \emph{V}=$(D_v,\sigma)$ where: 
(i) $D_v$ is a view of \emph{D} computed by 
eliminating inaccessible element types\footnote{An element type \emph{B} is inaccessible if for each parent type \emph{A}, 
either (i) the annotation \texttt{ann}(\emph{A,B})=\emph{N} 
exists or (ii) \emph{A} is an inaccessible element type.} from \emph{D}, according to annotations given in \emph{S};
(ii) $\sigma$ is a function used to extract accessible data in such a way that for each 
pair of types \emph{A} and \emph{B} where \emph{B} occurs in \emph{P(A)} in $D_v$, the $\sigma(A,B)$ is an XPath query 
(described in fragment $\mathcal{X}$) defining paths 
to reach element nodes \emph{B} from an element node \emph{A} in the original document \emph{T} conforms to \emph{D}. 
It should be noted that function  $\sigma$ is hidden from the users.
A security view \emph{V}=$(D_v,\sigma)$ is said \emph{recursive} if its $D_v$ is recursive.\newline

\begin{figure}[t!]
\begin{center}
\includegraphics[width=5cm]{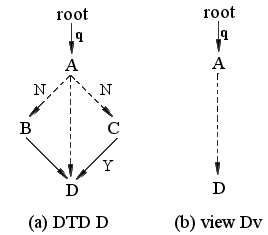}
\caption{Simple non-recursive DTD.}
\label{DTDExample}
\end{center}
\end{figure}

\noindent \textbf{Example 2.1} 
Consider the DTD \emph{D} depicted in Figure \ref{DTDExample}(a) where the labels of the edges represent the following access specification: 
\texttt{ann}(\emph{root,A})=[\emph{q}], \texttt{ann}(\emph{A,B})=\emph{N}, \texttt{ann}(\emph{A,C})=\emph{N}, and \texttt{ann}(\emph{C,D})=\emph{Y}.
We define the security view $V$=($D_v$,$\sigma$) as follows:\footnote{Note that $\epsilon$ denotes the empty path.}

\begin{algorithmic}
\STATE $D_v:$ $root\rightarrow A$, $A\rightarrow D|\epsilon$, $D\rightarrow \epsilon$.
\STATE \textbf{root $\rightarrow$ A}: we have \emph{$\sigma$(root,A)}=\emph{A}[\emph{q}].
\STATE \textbf{A $\rightarrow$ D|$\epsilon$}: we get \emph{$\sigma$(A,D)}=\emph{(C $\cup$ $\epsilon$)/D}.
\end{algorithmic}
\noindent Figure \ref{DTDExample}(b) depicts the resulting view $D_v$.\hfill\(\square\)

\subsection{DTD Recursion Problem}
Most existing approaches \cite{ref4,ref6,ref7} for securing access to XML documents are based on the notion of security view. 
Given a security view \emph{V}=$(D_v,\sigma)$, the query rewriting principle is 
applied to translate each XPath query \emph{p} over $D_v$ to another one $p_t$ over the original DTD \emph{D}, such that for any instance \emph{T} of \emph{D} 
($T_v$ of $D_v$ resp.), $p_t$ over \emph{T} yields the same answer as 
\emph{p} over $D_v$ (i.e. $p_t$(\emph{T}) $=$ $p$($T_v$))\footnote{We denote by \emph{p(T)} the result of evaluating query \emph{p} over document \emph{T}.}. Thanks to query rewriting we do not need to materialize view $T_v$ and its major problem namely
the view maintenance.
However, only non-recursive DTDs are considered. 
The security view as specified before cannot be applied in the case of 
recursive DTDs. To illustrate this problem we give the following example:\newline

\begin{figure}[t!]
\begin{center}
\includegraphics[width=6cm]{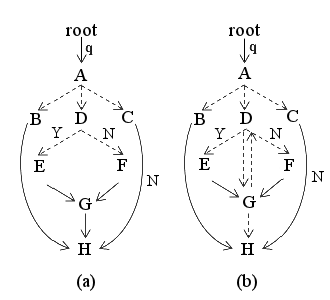}
\caption{DTD Recursion Problems.}
\label{DTDRecursion}
\end{center}
\end{figure}

\noindent \textbf{Example 2.2}  For the query \texttt{$\downarrow^{+}$::$H$} over the DTD given in Figure \ref{DTDRecursion}(a), 
we should enumerate all the paths from the \emph{root} which give an accessible element \emph{H} (as done in \cite{ref4}): 
\texttt{/$root$/$A$[$q$]/($B$ $\cup$ $D$/$E$/$G$)/$H$}. 
However, the task is complicated in the case of recursion. With the same query over the DTD in Figure \ref{DTDRecursion}(b), 
the function $\sigma$ used to extract accessible data, 
cannot be defined, e.g. $\sigma(D,E)$ can be $E$, $F/G/D/E$, or $F/G/D/F/G/D/E$ etc. Then $\sigma(D,E)$ leads to infinitely many paths 
and cannot be defined in $\mathcal{X}$. Moreover, the rewriting of the query \texttt{$\downarrow^{+}$::$H$} is equivalent 
to the following regular expression:

\begin{algorithmic}
\STATE /$root$/$A$[$q$]/$B$/$H$ $\cup$ /$root$/$A$[$q$]/$q_{1}$/($q_{1}$)*/$H$ $\cup$
\STATE /$root$/$A$[$q$]/($q_{2}$)*/\emph{D}/$E$/$G$/(\emph{D}/$G$)*/$H$
\end{algorithmic}

\noindent where:
\texttt{$q_{1}$ = \emph{D}/$G$ $\cup$ \emph{D}/$E$/$G$}, and \texttt{$q_{2}$ = \emph{D}/$G$ $\cup$ \emph{D}/$E$/$G$ $\cup$ \emph{D}/$F$/$G$}.\hfill\(\square\)\newline

Since the \emph{Kleene Star} (denoted by \texttt{*}) is not part of the standard XPath and cannot be expressed
as outlined in~\cite{ExpressivityXPath,ComplexityXPath20}, the rewriting 
of XPath queries is not always possible. We refer to this problem as the \emph{non-closure} of XPath fragment under the query rewriting. 
The \emph{closure} property is defined as follows:

\begin{definition}\label{closure}
A class $C$ of XPath queries is \emph{closed} under query rewriting if there is a function 
\emph{\texttt{Rewrite}}: $C\rightarrow C$ that, for any security view $V$=$(D_v,\sigma)$ and any query $Q$ in $C$ over $D_v$, 
computes $Q_t$=\emph{\texttt{Rewrite}($Q$)} in $C$ such that 
for any instance $T$ conforms to the original DTD $D$ and its view $T_v$ w.r.t $V$, we have \emph{$Q$($T_v$) = $Q_t$($T$)}.\hfill\(\square\)
\end{definition}

It has been shown in \cite{ref5} that the \emph{downward} class (i.e. fragment $\mathcal{X}$) 
of XPath queries is not \emph{closed} under query rewriting.

\begin{theorem}
For recursive XML security views, fragment $\mathcal{X}$ is not closed under query rewriting \cite{ref5}.\hfill\(\square\)
\end{theorem}

\subsection{Our Proposed Solution}
We show that the expressive power of the standard XPath \cite{XPath20} is sufficient to overcome the query rewriting problem over recursive views. 
We propose to redefine  
function \texttt{Rewrite} given in Definition~\ref{closure} into \texttt{Rewrite}: $C_1\rightarrow C_2$ where $C_2$ is the fragment $C_1$ extended by adding some 
axes and operators. Using this extension,
for any access specification \emph{S}=$(D,\texttt{ann})$ and any query \emph{Q} in $C_1$ over $D_v$ (view of \emph{D} computed w.r.t \emph{S}), 
we can compute $Q_t = \texttt{Rewrite}(Q)$ in $C_2$ such that for any instance \emph{T} of \emph{D} and its view $T_v$ w.r.t 
\emph{S}, we have $Q(T_v)=Q_t(T)$.\newline

The input fragment $C_1$ in our case is fragment $\mathcal{X}$ (namely the downward class) defined in Section \ref{XPathQueries}, which is used only to formulate user queries 
and to define access specifications, while $C_2$ is an extended fragment of $\mathcal{X}$ defined as follows:\newline

\begin{alltt}
 \texttt{path}    :=  \texttt{axis}::\texttt{label} | \texttt{path}'['\texttt{qual}']' | \texttt{path}'['n']'
            | \texttt{path}'/'\texttt{path} | \texttt{path}'\(\cup\)'\texttt{path}
 \texttt{qual}    :=  \texttt{path} | \texttt{path} = \(c\) | \texttt{path} = \(\varepsilon\)::\texttt{label}
            | \texttt{qual} \(and\) \texttt{qual} | \texttt{qual} \(or\) \texttt{qual} 
            | \(not\) \texttt{qual} | '('\texttt{qual}')'
 \texttt{axis}    :=  \(\varepsilon\) | \(\downarrow\) | \(\uparrow\) | \(\downarrow\sp{+}\) | \(\uparrow\sp{+}\) | \(\uparrow\sp{*}\)
\end{alltt}

\noindent we enrich $\mathcal{X}$ by the \emph{self}-axis ($\varepsilon$), the upward-axes \emph{parent} ($\uparrow$), \emph{ancestor} ($\uparrow^{+}$), and 
\emph{ancestor}-\emph{or}-\emph{self} ($\uparrow^{*}$), the \emph{position} and the \emph{node comparison} predicates.
The position predicate, defined with \texttt{[n]}(\emph{n $\in$ N}), is used to return the $n^{th}$ node from an ordered set of nodes. For instance, since we model 
the XML document with an ordered tree, the query 
$\downarrow::*[1]$ at an element node \emph{n} returns its first child element, while $\uparrow^{+}::*[B='topo'][1]$ returns its first ancestor element 
which has an child element \emph{B} with text content \emph{'topo'}. The \emph{node comparison} predicate \texttt{[$target_{1}$=$target_{2}$]} is true only if 
the evaluation of the right and left sides result in exactly the same 
single node. For example the predicate in the following query $\downarrow^{+}::A\downarrow::/B[\uparrow$::\texttt{*} = $\uparrow^{+}$::\texttt{*}$[1]]$ is valid for any \emph{B} element child of an \emph{A} element.

We summarize the augmented fragment of $\mathcal{X}$ by the following subsets $\mathcal{X}^{\Uparrow}$ ($\mathcal{X}$ with \emph{self} and \emph{upward}-axes), 
$\mathcal{X}^{\Uparrow}_{[n]}$ ($\mathcal{X}^{\Uparrow}$ with \emph{position} predicate), and the final fragment $\mathcal{X}^{\Uparrow}_{[n,=]}$ 
($\mathcal{X}^{\Uparrow}_{[n]}$ with \emph{node comparison} predicate).

It should be noted that for a given query \emph{Q}, a rewriting technique must ensure the following conditions: 
(i) each subquery\footnote{For example, the query $\downarrow^{+}$::$B$[$\downarrow$::$C$] contains two
subqueries $\downarrow^{+}$::$B$ and $\downarrow$::$C$
with a parent/child relation defined between 
\emph{C} and \emph{B}, and an ancestor/descendant relation defined between \emph{B} and the node context at which the query is posed.}
of \emph{Q} refers only to accessible element nodes; and (ii) each relationship defined between two subqueries of \emph{Q} is respected. For instance, 
the query $\downarrow^{+}$::\emph{D}/$\downarrow$::\texttt{*} over the access specification depicted in Figure \ref{DTDRecursion}(b) must returns only accessible 
element nodes of type \emph{G} or \emph{E} which have an accessible \emph{D} element as parent.

We define the accessibility problem as: ``\emph{When does an element node of a given type is accessible}?''. 
It is clear that the function $\sigma$ cannot solve this problem because of infinitely many possible paths involved by
recursive views (see Example 2.2).
We show in the next that 
the accessibility of a given element node w.r.t a given recursive view cannot be defined in the fragment $\mathcal{X}$ (even in $\mathcal{X}^{\Uparrow}$). 
We investigate the use of the augmented fragment $\mathcal{X}^{\Uparrow}_{[n]}$ to solve the accessibility problem in particular, and 
the fragment $\mathcal{X}^{\Uparrow}_{[n,=]}$ as a solution to avoid the non-closure of XPath fragment $\mathcal{X}$ in general.

\subsection{Notations}\label{NotationSection}
Given an access specification \emph{S}=$(D,\texttt{ann})$, and a document \emph{T} conforms to \emph{D}.
We define two predicates $\mathcal{A}_1^{acc}$ and $\mathcal{A}_2^{acc}$ as 
follows\footnote{Note that $A.\sigma(A',A)$ gives $A[Q]$ if \texttt{ann}($A'$,\emph{A})=[$Q$] and $A$ otherwise.}:

\begin{itemize}
\item[] \texttt{$\mathcal{A}_1^{acc}$ := $\uparrow^{*}$::*[$\varepsilon$::$root$ $\bigvee_{ann(A',A)\in ann}$ $\varepsilon$::$A$/$\uparrow$::$A'$][1]}\newline
\hspace*{15mm} \texttt{[$\varepsilon$::$root$ $\bigvee_{(ann(A',A)=Y|[Q])\in ann}$ $\varepsilon$::$A.\sigma(A',A)$/$\uparrow$::$A'$]}
\item[] \texttt{$\mathcal{A}_2^{acc}$ := $\bigwedge_{(ann(A',A)=[Q])\in ann}$ not ($\uparrow^{+}$::$A$[not (\emph{Q})]/$\uparrow$::$A'$)}
\end{itemize}

\noindent Where $\bigwedge$ and $\bigvee$ denote \emph{conjunction} and \emph{disjunction} respectively. The predicate $\mathcal{A}_1^{acc}$ has the form \texttt{$\uparrow^{*}$::*[$qual_1$][1][$qual_2$]}. 
Applying \texttt{$\uparrow^{*}$::*[$qual_1$]} on an element node \emph{n} of \emph{T} returns an ordered 
set $\mathcal{S}$ of element nodes (\emph{n} and/or some of its ancestor elements) such that for each one an annotation is defined. Thus, with 
$\mathcal{S}$[$qual_2$] ($n\vDash \mathcal{A}^{acc}_1$) we ensure that the first element node in $\mathcal{S}$ is concerned by a valid annotation.
With the second predicate, we use $n\vDash \mathcal{A}^{acc}_2$ to ensure that all qualifiers defined over ancestor elements of \emph{n} are valid (we discuss this restriction in Section \ref{AccessSpecificationRevision}).
These predicates are ``powerful tools'' to solve the accessibility problem as we will see in the next section.\newline

Since the $\sigma$ function is not computable in case of recursion, the parent/child relation defined between two element 
types in the query (e.g. query $\downarrow^{+}$::$A$/$\downarrow$::$B$ defines parent/child relation between \emph{A} and \emph{B}) cannot be rewritten in $\mathcal{X}$.
Accordingly, we define the two predicates \texttt{$\mathcal{A}^{+}$} and \texttt{$\mathcal{A}^{B}$}
to rewrite parent/child relation:

\begin{itemize}
\item[]\texttt{$\mathcal{A}^{+}$ := $\uparrow^{+}$::*[$\mathcal{A}_1^{acc}$]} 
\item[]\texttt{$\mathcal{A}^{B}$ := $\mathcal{A}^{+}$[1]/$\varepsilon$::$B$}
\end{itemize}

For an element node \emph{n} in \emph{T}, \emph{n}\textlbrackdbl $\mathcal{A}^{+}$ \textrbrackdbl$\;$ returns the set of 
all accessible ancestor elements of \emph{n}. The element node \emph{n} has an accessible \emph{B} element as parent if and only if $n\vDash \mathcal{A}^{B}$.

We use these four predicates throughout the paper to formalize our solution.

\section{Access Control with Recursive DTDs}\label{Sect3}
\begin{figure}[t!]
\begin{center}
\includegraphics[width=8cm,height=5cm]{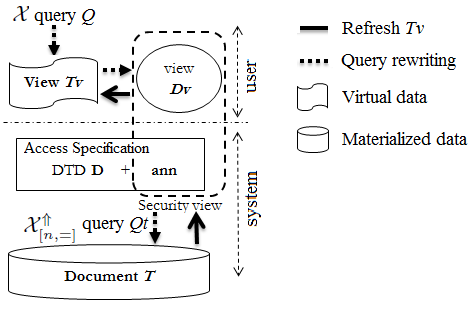}
\caption{XML Access Control Framework.}
\label{Framework}
\end{center}
\end{figure}

Our access control framework is presented in Figure \ref{Framework}. 
For each class of users, the administrator defines an access specification \emph{S}=$(D,\texttt{ann})$ over the DTD \emph{D}. The DTD view $D_v$ is derived first 
and given to the users to formulate their queries.
For each instance \emph{T} of \emph{D}, we compute a virtual\footnote{The views of \emph{T} are never materialized.} 
view $T_v$ of \emph{T} to show only accessible data. Each $\mathcal{X}$ query \emph{Q} over $T_v$ is efficiently rewritten, 
using the security view \emph{V} (defined below in Section \ref{sec:rec_views}), 
to an equivalent $\mathcal{X}^{\Uparrow}_{[n,=]}$ query $Q_t$ over \emph{T}, in order to return only accessible data.

\begin{figure}[t!]
\textbf{Algorithm:} \textit{DeriveView}\newline     
\begin{algo}
\begin{footnotesize}
\SetKwInOut{Input}{input}
\SetKwInOut{Output}{output}
\Input{an access specification \emph{S}=$(D,ann)$ with \emph{D}=$(Ele,P,root)$.}
\Output{a DTD view $D_v$.}
\texttt{\{$Ele_v$,$P_v$\} := \{$\phi$,$\phi$\}}\;
\texttt{Exp($root$,$true$)}\;
\ForEach{\texttt{$element$ $type$ $A\in D$}}{
\If{\texttt{Parsed($A$,$true$)$\neq \phi$}}{
\texttt{$Ele_v$ := $Ele_v$ $\cup$ $A$}\;
\texttt{$P_v$($A$) := Parsed($A$,$true$)}\;
}
}
\texttt{$D_v$ := $(Ele_v,P_v,root)$}\;
\Return $D_v$;\newline\newline\textbf{Procedure:} \textit{Exp}($A$, $access$)\newline\textbf{Input}: an element type \emph{A}, inherited accessibility $access$.\newline\textbf{Output}: content model of \emph{A}.\\
\setcounter{AlgoLine}{0}

\If{\texttt{Parsed($A$,$access$)$\neq null$}}{
\Return \texttt{Parsed($A$,$access$)}\;
}
\texttt{$exp$ := $\phi$}\;
\uCase{\texttt{\emph{P(A)} is $str$ or $\epsilon$}}{
\If{\texttt{$access$}}{
\texttt{$exp$ := $str$ ($\epsilon$ resp)}\;
}
}\Case{\texttt{\emph{P(A)} is $G_1$ $op_A$...$op_A$ $G_n$}}{
\tcp{$op_A$ is "|" or ","}
\ForEach{\texttt{subexpression $G_i$}}{
\uIf(\tcp*[h]{$case$ $of$ $single$ $element$ $type$}){\texttt{$G_i$ = \emph{B}}}{
\uIf{\texttt{$ann(A,B)\notin ann$}}{
\uIf{\texttt{$access$}}{
\texttt{$exp$ := $exp$ $op_A$ $B$; Exp($B$,$true$)}\;
}\ElseIf{\texttt{Exp($B$,$false$)$\neq \phi$}}{
\texttt{$exp$ := $exp$ $op_A$ Exp($B$,$false$)}\;
}
}\uElseIf{\texttt{$ann(A,B)=Y$}}{
\texttt{$exp$ := $exp$ $op_A$ $B$; Exp($B$,$true$)}\;
}\uElseIf{\texttt{$ann(A,B)=N$ and Exp($B$,$false$)$\neq \phi$}}{
\texttt{$exp$ := $exp$ $op_A$ Exp($B$,$false$)}\;
}\Else(\tcc*[h]{$ann(A,B)=[Q]$}){
\texttt{$exp$ := $exp$ $op_A$ ($B$|$\epsilon$)}\;
\texttt{Exp($B$,$true$)}\;
}
}\uElseIf{\texttt{$G_i$ = $B*$}}{
\texttt{$similar$ $to$ $the$ $previous$ $case$ $except$ $that$ $in$ $exp$, $B$ $is$ $replaced$ $with$ $B$* ($also$ $for$ Exp($B$,$true$) $and$ Exp($B$,$false$))}\;
}\Else(\tcp*[h]{$G_i$ $is$ $composition$ $of$ $element$ $types$/$subexpressions$}){
\texttt{$define$ $A'$ $in$ \emph{D} $as$ $temporary$ $element$ $type$}\;
\texttt{$define$ $content$ $model$ $P(A'):= G_i$}\;
\If{\texttt{Exp($A'$,$access$)$\neq \phi$}}{\texttt{$exp$ := $exp$ $op_A$ Exp($A'$,$access$)}\;}
\texttt{$delete$ $A'$ $from$ \emph{D} $and$ $the$ Parsed($A'$,$access$) $entry$}\;
}
}
}

\lIf{($exp=\phi$ and $access$)}{$exp$ := $\epsilon$\;}
\texttt{Parsed($A$,$access$):=$exp$}\;

\Return $exp$\;

\end{footnotesize}
\end{algo}
\caption{DTD View Derivation Algorithm.}
\label{Derive}
\end{figure}

\subsection{Recursive Security Views}\label{sec:rec_views}
We redefine the security view over an access specification \emph{S}=$(D,\texttt{ann})$ to 
be \emph{V}=$(D_{v},\texttt{ann})$, where $D_v$ is the view of \emph{D}, computed by algorithm \texttt{DeriveView} illustrated in
Figure \ref{Derive}, and used by the users to formulate their queries. 

We use first a DTD parser \footnote{Available at: \url{http://www.rpbourret.com/dtdparser/index.htm}.} to explore 
the DTD \emph{D} into an expressive indexed structure in such a way, for each element type \emph{A} in \emph{D}, the set of its children types and descendant 
types are returned, also the content model \emph{P(A)} is represented as a tree where all sub-expressions composing \emph{P(A)} are detected as we 
explain in the next.

We define the recursive function \texttt{Exp}(\emph{A,access}) that, according to a given access specification 
$S$=($D$,\texttt{ann}), extracts 
the content model for the element type \emph{A} and for all its descendant types in \emph{D}. 
For each element type \emph{A} parsed by \texttt{Exp}, we eliminate all inaccessible subelement types $B_i$ of \emph{A} 
($\texttt{if ann}(A,B_i)$=$N$ exists) to compute the new content model 
$P_v(A)$. The value \emph{access} represents the inherited accessibility. 
For a given content model "$A\longrightarrow B_1,(B_2|B_3)$", we refer 
by $G_1$ $op_A$ $G_2$ to the sub-expressions $B_1$ and $B_2|B_3$ respectively, separated by $op_A=","$.
The list \texttt{Parsed} is used (i) to store the extracted content model of each 
element type in \emph{D}; and (ii) to avoid more than one parsing of the same element type. 
By invoking \texttt{Exp}(\emph{root,true}) in algorithm \texttt{DeriveView}, the content models of 
all element types of \emph{D} are computed. The value of \texttt{Parsed}(\emph{A,true})=$\phi$ indicates that the element type \emph{A} is not accessible, while the value \texttt{Parsed}(\emph{A,true})=$\epsilon$ indicates that 
the content model of \emph{A} is an empty word. The output of algorithm \texttt{DeriveView} is a DTD view $D_v=(Ele_v,P_v,root)$
where $Ele_v$ is computed by eliminating inaccessible element types from \emph{D}, and $P_v$ 
returns the content model of each (accessible) element type in $D_v$.

The complexity of our DTD view derivation algorithm, \texttt{DeriveView}, is given by the following theorem:

\begin{theorem}
Let $S$=$(D,\emph{\texttt{ann}})$ be an access specification, and $P'$ be the largest production in $D$, then the view $D_v$ 
of $D$ can be derived w.r.t $S$ in at most $O(|D|*|P'|)$ time.\hfill\(\square\)
\end{theorem}

\textsc{Proof.} For an element type \emph{A} in \emph{D}, we denote by $|P(A)|$ the number of all 
subelement types and operators (``,'' or ``|'') defining \emph{P(A)}. 
The procedure \texttt{Exp} of Figure \ref{Derive} works over the hierarchical, 
parse-tree representation of the regular expression \emph{P(A)}. This tree is given using the DTD parser cited above, where its intermediate nodes 
represent operators and the leaves are the subelement types of \emph{A}. Each operator links 
two or more element types/sub-expressions. To compute the new content model of \emph{A}, we parse all the element types $B_i$ (i.e. the leaves) of the \emph{P(A)} tree to eliminate 
each inaccessible $B_i$ (i.e. \texttt{ann}$(A,B_i$)=$N$ exists). Next, each node operator with no children nodes is eliminated. Finally, the new resulting
tree  is translated into a regular expression which represents the content model $P_v(A)$. Thus, these steps are done by parsing all the nodes of the \emph{P(A)} tree in $O(|P(A)|)$ time. If we consider that $P'$ is the largest 
production in \emph{D} then $O(|P(A)|)$ is bounded by $O(|P'|)$ and the content models of all element types 
of \emph{D} are computed in at most $O(|D|*|P'|)$ time.\hfill\(\square\)\newline

\subsection{Accessibility}
Now we define the element node accessibility based on the use of recursive views:

\begin{definition}\label{accessibility}
Given a security view \emph{V}=$(D_{v},\emph{\texttt{ann}})$ and a document \emph{T} conforms to the original DTD \emph{D}, then an element node \emph{n} on \emph{T} of type 
\emph{B} with parent node of type \emph{A} is accessible (shown in the view $T_v$ of \emph{T}), if and only if the following conditions hold:

\begin{itemize}
\item[\textit{i})] The element node \emph{n} is concerned by a valid annotation, or 
$\emph{\texttt{ann}}(A,B)$ does not exist and there is an annotation defined over ancestor element $n'$ of \emph{n} where:
$n'$ is the first ancestor element of \emph{n} concerned by an annotation, and this annotation is valid at $n'$.
 \item[\textit{ii})] For each ancestor element $n'$ of \emph{n} concerned by an annotation with value \emph{[$Q'$]}, $n'\vDash Q'$ must be verified.\hfill\(\square\)
\end{itemize}
\end{definition}

\begin{figure}[t!]
\begin{center}
\includegraphics[width=5cm,height=5cm]{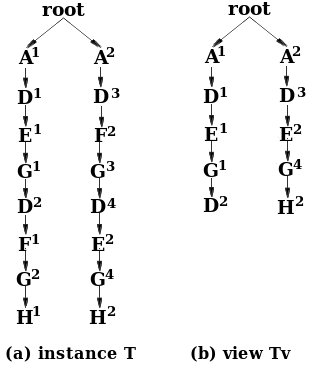}
\caption{Example of Instance view.}
\label{ViewInstance}
\end{center}
\end{figure}

\noindent \textbf{Example 3.1} Consider the DTD depicted in Figure \ref{DTDRecursion}(b) where annotation 
\texttt{q} is $\downarrow$::\emph{D}. 
For an element node \emph{n} of type \emph{H} within the instantiation of this DTD, 
if its parent element is of type \emph{C} then 
\emph{n} is not accessible. Otherwise, the first ancestor element of \emph{n} which is concerned by an annotation can be either of type \emph{F} (i.e. \texttt{ann}(\emph{D,F})=\emph{N}), 
of type \emph{E} (i.e. \texttt{ann}(\emph{D,E})=\emph{Y}), or of type \emph{A} (i.e. \texttt{ann}(\emph{root,A})=[\emph{q}]). This means that \emph{n} may be accessible if its first 
ancestor element is of type \emph{E} or \emph{A}, and it has no ancestor element $n'$ of type \emph{A} with $n'\nvDash q$.

\textit{Note that the element node accessibility over recursive XML views cannot be defined in $\mathcal{X}^{\Uparrow}$}. 
We consider the access specification \emph{S}=$(D,\texttt{ann})$ composed by the DTD of Figure \ref{DTDRecursion}(b) and the annotations depicted in 
the edges. Figure \ref{ViewInstance} represents (a) an instance \emph{T} of \emph{D} and (b) its view $T_v$ computed according to \emph{S}.
The query $\downarrow^{+}::H$ over \emph{T} must be rewritten to return only the node $H^{2}$, which is accessible w.r.t \emph{S}. 
However $\downarrow^{+}::H[\uparrow^{+}::E$ $or$ $\uparrow^{+}::A[q]]$ 
returns both the nodes $H^{1}$ and $H^{2}$, and $\downarrow^{+}::H[(\uparrow^{+}::E$ $or$ $\uparrow^{+}::A[q])$ $and$ $not$ $(\uparrow^{+}::F)]$ 
rejects the accessible node $H^{2}$ shown in $T_v$.\hfill\(\square\)\newline

We use the predicates $\mathcal{A}_1^{acc}$ and $\mathcal{A}_2^{acc}$ defined in Section \ref{NotationSection} to satisfy the 
accessibility conditions \textit{(\textbf{i})} and \textit{(\textbf{ii})} respectively of Definition \ref{accessibility}.

\begin{definition}\label{AccPred}
For any security view \emph{V}=$(D_v,\emph{\texttt{ann}})$ and any instance \emph{T} conforms to the original DTD \emph{D}, 
we define the accessibility predicate $\mathcal{A}^{acc}$ which refers to an 
$\mathcal{X}^{\Uparrow}_{[n]}$ qualifier such that, an element node \emph{n} on \emph{T} is accessible iff 
$n\vDash\mathcal{A}^{acc}$, with $\mathcal{A}^{acc}$ $:=$ $\mathcal{A}_1^{acc} \wedge \mathcal{A}_2^{acc}$.\hfill\(\square\)
\end{definition}

For an element type \emph{B} in a DTD \emph{D}, $\downarrow^{+}$::$B$[$\mathcal{A}^{acc}$] stands for all accessible \emph{B} elements in an instance of \emph{D}.\newline

\noindent \textbf{Example 3.2} Consider the access specification depicted in Figure \ref{DTDRecursion}(b), 
the predicates $\mathcal{A}_1^{acc}$ and $\mathcal{A}_2^{acc}$ are defined as follows:

\begin{itemize}
 \item[] \texttt{$\mathcal{A}_1^{acc}$ := $\uparrow^{*}$::*[$\varepsilon$::$root$ $\vee$ $\varepsilon$::$A$/$\uparrow$::$root$ $\vee$ $\varepsilon$::$E$/$\uparrow$::\emph{D} $\vee$ $\varepsilon$::$F$/$\uparrow$::\emph{D}}\newline
\texttt{$\vee$ $\varepsilon$::$H$/$\uparrow$::$C$][1][$\varepsilon$::$root$ $\vee$ $\varepsilon$::$A$[$q$]/$\uparrow$::$root$ $\vee$ $\varepsilon$::$E$/$\uparrow$::\emph{D}]}
 \item[] \texttt{$\mathcal{A}_2^{acc}$ := not ($\uparrow^{+}$::$A$[not ($q$)]/$\uparrow::root$)}
\end{itemize}

\noindent Consider the element node $H^{1}$ of the XML document illustrated in Figure \ref{ViewInstance}(a). Then, 
$H^{1}$\textlbrackdbl$\uparrow^{*}$::*[$\varepsilon$::$A$/$\uparrow$::$root$ $\vee$ $\varepsilon$::$E$/$\uparrow$::\emph{D} $\vee$ $\varepsilon$::$F$/$\uparrow$::\emph{D} $\vee$ $\varepsilon$::$H$/$\uparrow$::$C$ $\vee$ $\varepsilon$::$root$]\textrbrackdbl$\;$ returns the set 
$\mathcal{S}$=\{$F^{1}$,$E^{1}$,$A^{1}$,$root$\} of ordered element nodes (element node $H^{1}$ and/or some of its ancestor elements) where for each one an 
annotation exists (e.g. \texttt{ann}(\emph{D,F})=\emph{N} for element node $F^{1}$). Note that $\mathcal{S}$[1] returns the ancestor element $F^{1}$ and the final 
predicate 
$\mathcal{A}^{acc}_1$ over the element node $H^{1}$ of Figure \ref{ViewInstance}(a) is not satisfied 
(i.e. $H^{1}\nvDash \mathcal{A}^{acc}_1$) since the first ancestor element concerned by an annotation in $\mathcal{S}$ is not accessible 
($F^{1}$'s annotation is not valid).
The query $\downarrow^{+}$::$H$[$\mathcal{A}^{acc}$] over the instance \emph{T} of Figure \ref{ViewInstance}(a) 
returns only the accessible element $H^{2}$ (shown in the view $T_v$ of Figure \ref{ViewInstance}(b)).\hfill\(\square\)\newline

\textit{Property 1.} For any security view \emph{V}=$(D_{v},\texttt{ann})$, the accessibility predicate $\mathcal{A}^{acc}$ can be constructed in 
$O(|\texttt{ann}|)$ time.\hfill\(\square\)
\medskip

\textsc{Proof.} For any security view \emph{V}=$(D_{v},\texttt{ann})$, the construction of $\mathcal{A}_1^{acc}$ and $\mathcal{A}_2^{acc}$ depends only 
on the parsing of all annotations \emph{\texttt{ann}} of \emph{V} which is done in $O(|\texttt{ann}|)$ time.\hfill\(\square\)

\section{Query Rewriting over Recursive XML Views}\label{Sect4}
In this section we describe our XPath-based query rewriting algorithm.
Given an access specification \emph{S}=$(D,\texttt{ann})$, the security view \emph{V}=$(D_v,\texttt{ann})$ of \emph{S}, an instance \emph{T} conforms to \emph{D}, 
and its virtual view $T_v$ computed w.r.t \emph{V}.
Then, for any query \emph{p} over $T_v$, the goal of query rewriting is to find a rewriting function that we define as:\newline

\begin{algorithmic}
\STATE $\mathcal{X}$ $\longrightarrow$ $\mathcal{X}^{\Uparrow}_{[n,=]}$
\STATE $p$    $\longrightarrow$ \texttt{Rewrite}($p$) such that $p$($T_v$) = \texttt{Rewrite($p$)}(\emph{T})\newline
\end{algorithmic}

\noindent Our rewriting function \texttt{Rewrite} ensures that only accessible element nodes are referred to by the subqueries of \emph{p}, which is ensured 
by the accessibility predicate of Definition \ref{AccPred}. Moreover, the relationships defined between each two subqueries of \emph{p} must be respected.\newline

\begin{figure}[t!]
\begin{center}
\includegraphics[width=7cm]{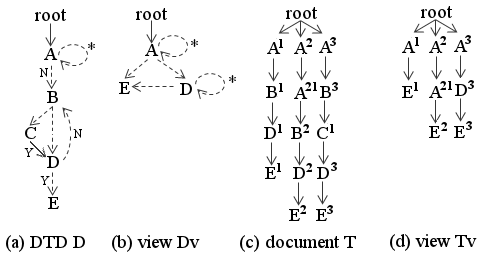}
\caption{Query Rewriting Problems.}
\label{AxesRewritingProblem}
\end{center}
\end{figure}

Notice that, for a given security view \emph{V}=$(D_v,\texttt{ann})$, we compute first the predicates $\mathcal{A}^{acc}$ and $\mathcal{A}^{+}$ w.r.t \emph{V} 
in $O(|\texttt{ann}|)$ time (Properties 1 and 2 resp.). 
Also, for each element type \emph{A} in $D_v$ we compute the lists of its children types and descendant types 
denoted by \texttt{Reach}($\downarrow$,\emph{A}) and \texttt{Reach}($\downarrow^{+}$,\emph{A}) respectively. 
Each list is computed in $O(|D_v|)$ time. The lists of all element types of $D_v$ are computed in $O(|D_v|^{2})$ time.
This preprocessing step is done only one time after the security view \emph{V} is defined and it provides performance gains during the query rewriting step.\newline

In this section, we consider the DTD view $D_{v}$ shown in Figure \ref{AxesRewritingProblem}(b) 
which represents the derivation of the DTD \emph{D} of Figure \ref{AxesRewritingProblem}(a) with respect to the access specification depicted in the edges. 
Figure \ref{AxesRewritingProblem}(c) represents a valid instance \emph{T} of \emph{D} and its derived view $T_{v}$ is depicted in Figure \ref{AxesRewritingProblem}(d).

\subsection{Queries Without Predicates}
\label{AxesRewriting}
For a DTD \emph{D}=(\emph{Ele,P,root}), we discuss the rewriting of queries without predicates with the form 
\texttt{$axis_{1}$::$E_{1}$/.../$axis_{n}$::$E_{n}$} where $E_i$ $\in$ \{\emph{Ele},\texttt{*}\} and $axis_i$ is an XPath axis in fragment $\mathcal{X}$.
Given the query \texttt{$\downarrow^{+}$::$E_i$/$\downarrow^{+}$::$E_j$}, it is clear that the rewritten query can be 
\texttt{$\downarrow^{+}$::$E_i$[$\mathcal{A}^{acc}$]/$\downarrow^{+}$::$E_j$[$\mathcal{A}^{acc}$]} to return accessible $E_j$ elements which have at 
least one accessible ancestor element of type $E_i$. However, it is not so simple in the case of \emph{child}-axis.\newline

\noindent \textbf{Example 4.1} The query \texttt{$\downarrow^{+}$::$A$/$\downarrow$::$E$} over the view $T_{v}$ of Figure \ref{AxesRewritingProblem}(d) 
returns \emph{E} elements having an \emph{A} element as parent. So the elements $E^{1}$ and $E^{2}$ are returned. Using the kleene closure, this query can be rewritten 
over the instance \emph{T} of Figure \ref{AxesRewritingProblem}(c) into: \texttt{$\downarrow^{+}$::$A$/$\downarrow$::$B$/$\downarrow$::\emph{D}/($\downarrow$::$B$/$\downarrow$::\emph{D})*/$\downarrow$::$E$}.
However, using the standard XPath, a cycle in the DTD cannot be replaced by \texttt{'$\downarrow^{+}$'}. For instance the query 
\texttt{$\downarrow^{+}$::$A$[$\mathcal{A}^{acc}$]/$\downarrow^{+}$::$E$[$\mathcal{A}^{acc}$]} returns the elements $E^{1}$, $E^{2}$, and $E^{3}$, while 
$E^{3}$ does not have a parent \emph{A}.\hfill\(\square\)\newline

We use the predicate $\mathcal{A}^{B}$ (\emph{B} can be any element type) defined in Section \ref{NotationSection} to rewrite the parent/child relation.
For instance, the query \texttt{$\downarrow^{+}$::$A$/$\downarrow$::$E$} of the previous example can be rewritten into \texttt{$\downarrow^{+}$::$E$[$\mathcal{A}^{acc}$][$\mathcal{A}^{A}$]} to 
return only accessible \emph{E} elements (verified with $\mathcal{A}^{acc}$) which have an accessible parent of type \emph{A} (verified with $\mathcal{A}^{A}$).

\noindent \textbf{Example 4.2} Given the access specification of Figure \ref{AxesRewritingProblem}(a), we define the predicate $\mathcal{A}^{+}$ as follows:

\begin{algorithmic}
\STATE \texttt{$\mathcal{A}^{+}$ := $\uparrow^{+}$::*[$\uparrow^{*}$::*[$\varepsilon$::$E$/$\uparrow$::\emph{D} $\vee$ $\varepsilon$::$B$/$\uparrow$::\emph{D} $\vee$ $\varepsilon$::\emph{D}/$\uparrow$::$C$ $\vee$}
\STATE \texttt{$\varepsilon$::$B$/$\uparrow$::$A$ $\vee$ $\varepsilon$::$root$][1][$\varepsilon$::$E$/$\uparrow$::\emph{D} $\vee$ $\varepsilon$::\emph{D}/$\uparrow$::$C$ $\vee$ $\varepsilon$::$root$]]}.
\end{algorithmic}

\noindent At element node $E^{3}$ of Figure \ref{AxesRewritingProblem}(d), $\mathcal{A}^{+}$ returns the set of its ordered accessible ancestor elements 
\{$D^{3}$,$A^{3}$,$root$\}, that we denote $E^{3}$\textlbrackdbl $\mathcal{A}^{+}$\textrbrackdbl. The predicate $\mathcal{A}^{A}$ (i.e. $\mathcal{A}^{+}$[1]/$\varepsilon$::$A$)
does not hold at element node $E^{3}$ (i.e. $E^{3}\nvDash \mathcal{A}^{A}$) since 
the first accessible ancestor element of $E^{3}$ is not of type 
\emph{A} (i.e. $\mathcal{A}^{+}$[1] at $E^{3}$ returns its ancestor element $D^{3}$). However, $\mathcal{A}^{A}$ holds at element nodes $E^{1}$ and $E^{2}$.\hfill\(\square\)\newline

\textit{Property 2.} For any security view \emph{V}=$(D_{v},\texttt{ann})$ and any element type \emph{B} in $D_v$,
the predicates $\mathcal{A}^{+}$ and $\mathcal{A}^{B}$
can be constructed in $O(|\texttt{ann}|)$ time.\hfill\(\square\)\newline

\textsc{Proof.} The same principle as the proof of Property 1.\hfill\(\square\)\newline

Finally, given a (recursive) security view \emph{V}=$(D_{v},\texttt{ann})$, 
we define the rewriting function 
\texttt{Rewrite : $\mathcal{X}\times Ele$ $\rightarrow$ $\chi^{\Uparrow}_{[n]}$} that we use to rewrite an $\mathcal{X}$ 
query $p$=\texttt{$p_1$/.../$p_n$} (where each subquery $p_i$ is given with $axis_i$::$E_{i}$) 
over a node context of type \emph{E} in $D_v$, to an equivalent one \texttt{Rewrite}(\emph{p,E}) defined in $\mathcal{X}^{\Uparrow}_{[n]}$ over the original DTD as:\newline

\begin{algorithmic}
\STATE \texttt{Rewrite($p$,$E$) := $\downarrow^{+}$::$E_{n}$[$\mathcal{A}^{acc}$][$prefix^{-1}$($p_{1}$/.../$p_{n}$)]}\newline
\end{algorithmic}

\noindent Where the qualifier \texttt{$prefix^{-1}$($p_{1}$/.../$p_{n}$)} is recursively defined over the descending list of subqueries of \emph{p}. 
For each subquery $p_{i}$, \texttt{$prefix^{-1}$($p_{1}$/.../$p_{i-1}$)} is already computed and used to compute \texttt{$prefix^{-1}$($p_{1}$/.../$p_{i}$)} 
as follows:

\begin{itemize}
\item $axis_{i}$ = $\downarrow$: $prefix^{-1}$($p_{1}$/.../$p_{i}$) := $\mathcal{A}^{E_{i-1}}$[$prefix^{-1}$($p_{1}$/.../$p_{i-1}$)]
\item $axis_{i}$ = $\downarrow^{+}$: $prefix^{-1}$($p_{1}$/.../$p_{i}$) := $\uparrow^{+}$::$E_{i-1}$[$\mathcal{A}^{acc}$][$prefix^{-1}$($p_{1}$/.../$p_{i-1}$)]
\end{itemize}

\noindent Recall that \emph{E} is the type of context node at which the query is evaluated. 
As a special case we have: $prefix^{-1}$($\downarrow$::$E_{1}$)=$\mathcal{A}^{E}$, and 
$prefix^{-1}$($\downarrow^{+}$::$E_{1}$)=$\uparrow^{+}$::$E$[$\mathcal{A}^{acc}$].\newline

\noindent \textbf{Example 4.3} Consider the query \texttt{$\downarrow$::$A$/$\downarrow$::$E$} over the node context \emph{root} of 
the view $T_{v}$ of Figure \ref{AxesRewritingProblem}(d). Using our algorithm \texttt{Rewrite} we obtain:\newline

\begin{algorithmic}
\STATE \texttt{Rewrite($\downarrow$::$A$/$\downarrow$::$E$, $root$) = $\downarrow^{+}$::$E$[$\mathcal{A}^{acc}$][$\mathcal{A}^{A}$[$\mathcal{A}^{root}$]]=}
\STATE \texttt{$\downarrow^{+}$::$E$[$\mathcal{A}^{acc}$][$\mathcal{A}^{+}$[1]/$\varepsilon$::$A$[$\mathcal{A}^{+}$[1]/$\varepsilon$::$root$]]}\newline
\end{algorithmic}

\noindent $\mathcal{A}^{+}$ is given in Example 4.2. The evaluation of $\mathcal{A}^{+}$ returns \{$A^{1}$,$root$\} at 
element node $E^{1}$, \{$A^{21}$,$A^{2}$,$root$\} at $E^{2}$, and 
\{$D^{3}$,$A^{3}$,$root$\} at $E^{3}$. With $\mathcal{A}^{A}$ we ensure that for an element node \emph{E} referred to by the query, its first accessible ancestor element is 
of type \emph{A} which is verified for $E^{1}$, $E^{2}$ and not for $E^{3}$ in $T_{v}$ of Figure \ref{AxesRewritingProblem}(d). Moreover, with 
$\mathcal{A}^{root}$ we ensure that the \emph{A} element returned by $\mathcal{A}^{A}$, which can be $A^{1}$ or $A^{21}$ for element nodes $E^{1}$ and $E^{2}$ 
respectively, must have \emph{root} as the first accessible ancestor which is verified 
only for $A^{1}$. Thus the query \texttt{Rewrite($\downarrow$::$A$/$\downarrow$::$E$, $root$)} at \emph{root} 
of Figure \ref{AxesRewritingProblem}(d) returns \{$E^{1}$\}.\hfill\(\square\)\newline

\begin{figure}[t!]
\textbf{Algorithm:} \textit{Rewrite}\newline           
\begin{algo}
\begin{footnotesize}
\SetKwInOut{Input}{input}
\SetKwInOut{Output}{output}
\Input{A query $p$, and an element type \emph{A} for which query rewriting is carried.}
\Output{a rewritten query $p_t$ w.r.t \emph{A}.\newline}
\If{\texttt{$p$ = $p_{1}\cup$...$\cup p_{n}$}}{
\texttt{\Return $\cup_{\leq i \leq n}$ Rewrite($p_{i}$,\emph{A})}\;
}
\texttt{reach} := $\{A\}$\;
\textit{compute the descending list $L$ of subqueries of $p$}\;
\textit{each subquery $p_{i}$ = $axis_{i}$::$E_{i}[f_{i}]$}\;
$p_t$ := $\epsilon$; $filters$ := $\epsilon$\;
\tcp{$compute$ $prefix^{-1}(p)$ $to$ $be$ $p_t$}
\ForEach{$p_{i}$ in the order of $L$}{
\tcp{$axes$ $rewriting$}

\uCase{$axis_{i}=\downarrow$}{
\uIf{\texttt{($filters$ = $\epsilon$)}}{
$p_t$ := $\mathcal{A}^{fs(reach,\varepsilon)}$[$p_t$]\;
}\Else{
$p_t$ := $\mathcal{A}^{fs(reach,\varepsilon)}$[$filters$][$p_t$]\;
}
}\Case{$axis_{i}=\downarrow^{+}$}{
\uIf{\texttt{($filters$ = $\epsilon$)}}{
$p_t$ := $fs(reach,\uparrow^{+})[\mathcal{A}^{acc}][p_t]$\;
}\Else{
$p_t$ := $fs(reach,\uparrow^{+})[\mathcal{A}^{acc}][filters][p_t]$\;
}
}
\tcp{$*$-$label$ $elimination$}
\uIf{($E_{i}=*$)}{
\texttt{reach} := $\cup_{E\in reach}$\texttt{Reach}($axis_{i},E$)\;
}\uElseIf{($\exists$ $E\in$ reach $s.t$ $E_{i} \in$ \texttt{Reach}($axis_{i},E$))}{
\texttt{reach} := \{$E_{i}$\}\;
}\Else{
\texttt{reach} := \{\}\;
}
\lIf{(\texttt{\emph{reach}}=\{\})}{
\Return $\phi$\;
}
\tcp{$rewriting$ $of$ $predicate$ $f_{i}$ $over$ reach $elements$}
$filters$ := \texttt{RW\_Pred($f_{i}$,reach)}\;
\uIf(\tcp*[h]{$invalid$ $predicate$}){\texttt{($filters$ = $false$)}}{\Return $\phi$\;}
\ElseIf(\tcp*[h]{$omitte$ $f_i$ $from$ $p_i$}){\texttt{($filters$ = $true$)}}{$filters$ := $\epsilon$\;}
}
\tcp{$rewritten$ $query$ $p_t$ $of$ $p$ $w.r.t$ $A$}
\uIf{\texttt{($filters$ = $\epsilon$)}}{
$p_t$ := $fs(reach,\downarrow^{+})$[$\mathcal{A}^{acc}$][$p_t$]\;
}\Else{
$p_t$ := $fs(reach,\downarrow^{+})$[$\mathcal{A}^{acc}$][$filters$][$p_t$]\;
}
\Return $p_t$;\newline
\end{footnotesize}
\end{algo}
\caption{Algorithm for XPath Queries Rewriting.}
\label{RWAxes}
\end{figure}

The detail of Function \texttt{Rewrite} is given in Figure \ref{RWAxes}. After computing the descending list \emph{L} 
of  \emph{p}'s subqueries, we parse them to generate the $prefix^{-1}$ of the rewritten query as explained above.
The node context \emph{A} can be initialized to \emph{root} of the DTD for rewriting \emph{p} over the entire document. 
If $p_i$ is $axis_i$::\texttt{*}, then the \texttt{*}-label is replaced by the set of 
children/descendant types of $E_{i-1}$ ($axis_i$ is $\downarrow$ or $\downarrow^{+}$ resp.). Then, the rewriting of $p_i$ over $p_{i-1}$ can result in a set of element types denoted by \texttt{reach}. 
Moreover, if $E_i\neq$\texttt{*}, then it must exist at least one element type \emph{E} in \texttt{reach} (result of the rewriting of $p_{i-1}$) 
where $E_i$ is a child type of \emph{E} if $axis_{i}=\downarrow$ or one of its descendant types if $axis_{i}=\downarrow^{+}$ (lines 18-23). 
If the rewriting of $p_i$ over $p_{i-1}$ stands for an empty set \texttt{reach} then the query is rejected (line 24).
By \emph{fs} we refer to the function fusion, e.g. $fs(\{E_{1},...,E_{n}\},\uparrow^{+})$ = \texttt{$(\uparrow^{+}$::$E_{1} \cup..\cup \uparrow^{+}$::$E_{n})$}.\newline

Function \texttt{RW\_Pred}, called in the algorithm \texttt{Rewrite}, represents the predicates rewriting, which is the subject of the next section.
As shown above, \emph{upward}-axes and the \emph{position} predicate are necessary for rewriting simple queries (without predicates). We prove below that 
fragment $\mathcal{X}^{\Uparrow}_{[n]}$ is not closed under query rewriting. Extending this fragment with the \emph{node comparaison} operator (which results in the final fragment $\mathcal{X}^{\Uparrow}_{[n,=]}$)
turns out sufficient to rewrite any query in $\mathcal{X}$.

\begin{theorem}\label{NoClosurePredicateRewriting} For recursive XML security views, the XPath fragment $\mathcal{X}^{\Uparrow}_{[n]}$ is not closed under 
query rewriting.\hfill\(\square\)\end{theorem}

\textsc{Proof.} (\textit{by contradiction}) 
We consider query with the 
form \texttt{$E$[$q$]} which represents the rewriting limitation of the XPath fragment $\mathcal{X}^{\Uparrow}_{[n]}$.
Assume that the query rewriting can be done in $\mathcal{X}^{\Uparrow}_{[n]}$.  
The query $\downarrow^{+}$::$A$[$\downarrow$::$E$] over the view instance $T_{v}$ depicted in Figure \ref{AxesRewritingProblem}(d) cannot be correctly rewritten, using the previous definition of algorithm \texttt{Rewrite}, 
into \texttt{Rewrite}($\downarrow^{+}$::$A$,$root$)[\texttt{Rewrite}($\downarrow$::$E$,\emph{A})]
equivalent to $\downarrow^{+}$::$A$[$\mathcal{A}^{acc}$][$\downarrow^{+}$::$E$[$\mathcal{A}^{acc}$][$\mathcal{A}^{A}$]]. Indeed, the resulting query returns 
\{$A^{1}$, $A^{2}$, $A^{21}$\}, but $A^{2}$ does not have an immediate child $E$. The limitation is due to the fact that 
predicate [$\downarrow$::$E$] must return all descendant elements \emph{E} having as the first accessible ancestor the node context \emph{A} at which the 
predicate is evaluated (i.e. the element node returned by $\downarrow$::$E$/$\mathcal{A}^{+}$[1] must be the same element node of type \emph{A} at which the predicate is evaluated). 
This cannot be expressed in $\mathcal{X}^{\Uparrow}_{[n]}$ and can be done only by introducing the 
node set comparison (e.g. $\mathcal{X}^{\Uparrow}_{[n,=]}$) as we will present in the following.\hfill\(\square\)

\subsection{Predicates Rewriting}
We explain the rewriting of predicates to complete the definition of our rewriting algorithm \texttt{Rewrite}.
For a given query $axis_1$::$E_1$[$q_1$]/.../$axis_n$::$E_n$[$q_n$], we rewrite each predicate $q_i$ over the element type $E_i$ at which $q_i$ is defined.
Given a security view \emph{V}=$(D_{v},ann)$ and a subquery \texttt{$E$[$q$]} (we take a simple predicate $q$=$q_{1}$/.../$q_{n}$ where \texttt{$q_{i}$=$axis_{i}$::$E_{i}$} for more comprehension). 
We define the function \texttt{RW\_Pred : $\mathcal{X}$ $\times$ \emph{Ele} $\rightarrow$ $\mathcal{X}^{\Uparrow}_{[n,=]}$} to rewrite the predicate 
\texttt{$q$} in $\mathcal{X}$ over element type $E$ in $D_{v}$, 
to an equivalent one \texttt{RW\_Pred}($q$, $E$) in $\mathcal{X}^{\Uparrow}_{[n,=]}$, recursively defined over the descending list of sub-predicates of $q$ as follows:

\begin{itemize}
\item \texttt{$axis_{i}=\downarrow$}: \texttt{RW\_Pred($q_{i}$/.../$q_{n}$,$E_{i-1}$) := }\newline
\texttt{$\downarrow^{+}$::$E_{i}$[$\mathcal{A}^{acc}$][RW\_Pred($q_{i+1}$/.../$q_{n}$,$E_{i}$)]/$\mathcal{A}^{+}$[1] = $\varepsilon$::$E_{i-1}$}
\item \texttt{$axis_{i}=\downarrow^{+}$}: \texttt{RW\_Pred($q_{i}$/.../$q_{n}$,$E_{i-1}$) := }\newline
\texttt{$\downarrow^{+}$::$E_{i}$[$\mathcal{A}^{acc}$][RW\_Pred($q_{i+1}$/.../$q_{n}$,$E_{i}$)]}
\end{itemize}

\noindent Given a query $axis_i$::$E_i$[$axis_{j}$::$E_{j}$=$"c"$] (text-content comparison), we have the following rewriting:

\begin{itemize}
\item[] \texttt{RW\_Pred($\downarrow$::$E_{j}$=$"c"$,$E_{i}$) := }
\texttt{$\downarrow^{+}$::$E_{j}$[$\mathcal{A}^{acc}$][$\varepsilon$::*=$"c"$]/$\mathcal{A}^{+}$[1] = $\varepsilon$::$E_{i}$}
\item[] \texttt{RW\_Pred($\downarrow^{+}$::$E_{j}$=$"c"$,$E_{i}$) :=}
\texttt{$\downarrow^{+}$::$E_{j}$[$\mathcal{A}^{acc}$][$\varepsilon$::*=$"c"$]}
\end{itemize}

\noindent \textbf{Example 4.4} The query \texttt{$\downarrow^{+}$::$A$[$\downarrow$::$E$]}, given in the proof of Theorem \ref{NoClosurePredicateRewriting}, is 
rewritten into:
\begin{algorithmic}
\STATE \texttt{Rewrite}($\downarrow^{+}$::$A$[$\downarrow$::$E$], $root$) = $\downarrow^{+}$::$A$[$\mathcal{A}^{acc}$][$\uparrow^{+}$::$root$[$\mathcal{A}^{acc}$]][\texttt{RW\_Pred}($\downarrow$::$E$, $A$)]
\STATE  $\equiv$ $\downarrow^{+}$::$A$[$\mathcal{A}^{acc}$][$\downarrow^{+}$::$E$[$\mathcal{A}^{acc}$]/$\mathcal{A}^{+}$[1]=$\varepsilon$::$A$]
\end{algorithmic}
\noindent where $\uparrow^{+}$::$root$[$\mathcal{A}^{acc}$] is omitted since $root$ is always accessible. The rewritten query over $T_v$ of Figure \ref{AxesRewritingProblem}(d), 
returns the element nodes $A^{1}$ and $A^{21}$.\hfill\(\square\)\newline

The detail of Function \texttt{RW\_Pred} is given in Figure \ref{RWPred}. We have seen in \texttt{Rewrite} algorithm 
that the rewriting of subquery \texttt{$axis_{i+1}$::*} over $E_{i}$ results in a set of element types (\texttt{reach}) reachable from $E_{i}$. 
Then, the predicate \emph{q} in 
\texttt{$axis_i$::$E_i$[$q$]} is rewritten over element type $E_i$ ($E_i\neq$\texttt{*}) as explained in the above definition of \texttt{RW\_Pred}. While the predicate $q$ in \texttt{$axis_i$::$E_i$/$axis_{i+1}$::*[$q$]} 
is rewritten over the set of element types resulting by the rewriting of \texttt{$axis_{i+1}$::*} over $E_i$. We denote this set by \emph{L}.
For a given predicate \texttt{$q_1$/.../$q_n$} over element type \emph{E}, \emph{L} :=\texttt{reach($q_1$,$\{E\}$)} denotes the result of rewriting sub-predicate 
$q_1$ over $E$ (element types reachable from $E$ with $q_1$), 
sub-predicate $q_2$ is rewritten over \emph{L} resulting in a new set \emph{L} :=\texttt{reach($q_2$,$L$)} (i.e. \emph{L} :=\texttt{reach($q_2$,reach($q_1$,$\{E\}$))}), and 
so on until rewriting $q_n$ over \emph{L}:=\texttt{reach($q_{n-1}$,$L$)}.
Each sub-predicate \texttt{$q_i$} can contain other sub-predicates (case of $axis_i$::$E_i$[$f_i$]). 
The \texttt{*}-labels in the sub-predicates are eliminated with the same principle explained in algorithm \texttt{Rewrite} using the precomputed lists of 
children and descendant types (\texttt{Reach}).
The rewriting result of predicate $q$ can be \emph{false} if some element types in $q$ are inaccessible (they do not appear in $D_{v}$) or some relationships are not respected 
(e.g. the rewriting of predicate $\downarrow$::$E_i$ over element type $E_{i-1}$ is false if $E_i$ is not a subelement type of $E_{i-1}$, 
such that $E_i\notin$ \texttt{Reach}($\downarrow$,$E_{i-1}$)). 
Rewriting result can be \emph{true} (the predicate is omitted from the query) in case of \emph{not(q)} with non valid $q$.\newline

\begin{figure}[t!]
\textbf{Algorithm:} \textit{RW\_Pred}\newline
\begin{algo}
\begin{footnotesize}
\SetKwInOut{Input}{input}\SetKwInOut{Output}{output}
\Input{a predicate $q$ and a list $L$ of element types.}
\Output{a rewritten predicate $q_t$ w.r.t element types in $L$.}
\texttt{$q_t$ := $false$}, $f'$ := $true$\;
\tcc{\textit{$content$-$test$ is optional, if it does not exists then [$\varepsilon$::$E$=$c$] below is omitted}}
\uIf(\tcp*[h]{[$f$] is optional}){\texttt{($q$ is a single predicate $axis$::$E$[$f$]$=c$)}}{
\uIf{\texttt{($E=*$)}}{
\texttt{reach($q$,$L$) := $\cup_{E'\in L}$\texttt{Reach}($axis_{i},vE'$)}\;
}\uElseIf{($\exists$ $E'\in$ $L$ $s.t$ $E \in$ \texttt{Reach}($axis_{i},E'$))}{
\texttt{reach($q$,$L$) := $\{E\}$}\;
}\Else{
\texttt{reach($q$,$L$) := $\{\}$}\;
}
\lIf{(\texttt{\emph{reach}}($q$,$L$)=\{\})}{
\Return $false$\;
}
\tcp*[h]{rewriting of $f'$ if [$f$] exists}\\
\texttt{$f'$ := \textit{RW\_Pred}($f$,$E$)}\;
\If{(\emph{[$f$]} exists and \texttt{$f'=false$})}{
\Return $false$\;
}
\tcp*[h]{[$f'$] $below$ $is$ $omitted$ $if$ $f'$=$true$}\\
\uIf{\texttt{($axis=\downarrow$)}}{
\texttt{$q_t$ := $fs(\downarrow^{+},reach(q,L))$[$\mathcal{A}^{acc}$][$f'$][$\varepsilon$::$E$=$c$]/}\texttt{$\mathcal{A}^{+}$[1]=$\varepsilon$::$*$}\;
}\Else{
\texttt{$q_t$ := $fs(\downarrow^{+},reach(q,L))$[$\mathcal{A}^{acc}$][$f'$][$\varepsilon$::$E$=$c$]}\;
}
}\uElseIf{\texttt{($q$ is $q_{f}$/$q_{r}$ where $q_{f}$ = $axis_1$::$E$[$f$] and $q_r$ is the remaining steps)}}{
\emph{rewrite $q_{f}$ as done in lines 3-16}\;
\texttt{$q'_{r}$ := \textit{RW\_Pred}($q_{r}$,reach($q_{f}$,$L$))}\;

\lIf{($q'_{r}$=$false$)}{\Return $false$\;}

\uIf{\texttt{($axis_{1}=\downarrow$)}}{
\texttt{$q_t$ := $fs(\downarrow^{+},reach(q_{f},L))$[$\mathcal{A}^{acc}$][$f'$]}\texttt{[$q'_{r}$]/$\mathcal{A}^{+}$[1]=$\varepsilon$::$*$}\;
}\Else{
\texttt{$q_t$ := $fs(\downarrow^{+},reach(q_{f},L))$[$\mathcal{A}^{acc}$][$f'$][$q'_{r}$]}\;
}
}\uElseIf{\texttt{($q$ is $q_{1}\wedge$...$\wedge q_{n}$)}}{
\uIf{\texttt{($\exists$ $q_{i}$ $s.t$ \textit{RW\_Pred}($q_{i}$,$L$)$=false$)}}{
\texttt{$q_t$ := $false$}\;
}\uElseIf{\texttt{(\textit{RW\_Pred}($q_{i}$,$L$)$=true$ $for$ $each$ $q_{i}$)}}{
\texttt{$q_t$ := $true$}\;
}\Else{
\texttt{$q_t$ := $\bigwedge_{RW\_Pred(q_{i},L)\neq true}$\textit{RW\_Pred}($q_{i}$,$L$)}\;
}
}\uElseIf{\texttt{($q$ is $q_{1} \vee$...$\vee q_{n}$ or $q_{1} \cup$...$\cup q_{n}$)}}{
\uIf{\texttt{($\exists$ $q_{i}$ $s.t$ \textit{RW\_Pred}($q_{i}$,$L$)$=true$)}}{
\texttt{$q_t$ := $true$}\;
}\Else{
\texttt{$q_t$ := $\bigvee_{RW\_Pred(q_{i},L)\neq false}$\textit{RW\_Pred}($q_{i}$,$L$)}\;
}
}\uElseIf{case of \texttt{not ($q$)}}{
\uIf{\texttt{\textit{RW\_Pred}($q$,$L$)$=false$}}{
\texttt{$q_t$ := $true$}\;
}\ElseIf{\texttt{\textit{RW\_Pred}($q$,$L$)$\neq true$}}{
\texttt{$q_t$ := not (\textit{RW\_Pred}($q$,$L$))}\;
}
}\ElseIf{case of \texttt{$\epsilon$}}{
\texttt{$q_t$ := $true$}\;
}
\Return $q_t$\;
\end{footnotesize}
\end{algo}
\caption{Predicate Rewriting.}
\label{RWPred}
\end{figure}

\noindent \textbf{Example 4.5} Consider the query $\downarrow^{+}$::$A$[$\downarrow$::\texttt{*}/$\downarrow$::\emph{D}] 
over the Figure \ref{AxesRewritingProblem}(b). Using our algorithm \texttt{RW\_Pred}, the predicate $\downarrow$::\texttt{*}/$\downarrow$::\emph{D} 
is rewritten over element type \emph{A} as follows:
\begin{algorithmic}
\STATE [\texttt{RW\_Pred}($\downarrow$::\texttt{*}/$\downarrow$::\emph{D},\emph{A})] = 
\STATE [($\downarrow^{+}$::$A\cup\downarrow^{+}$::$D\cup\downarrow^{+}$::$E$)[$\mathcal{A}^{acc}$][\texttt{RW\_Pred}($\downarrow$::\emph{D},\{$A$, \emph{D}, $E$\})]/$\mathcal{A}^{+}$[1]=$\varepsilon$::$A$] =
\STATE [($\downarrow^{+}$::$A\cup\downarrow^{+}$::$D\cup\downarrow^{+}$::$E$)[$\mathcal{A}^{acc}$][$\downarrow^{+}$::\emph{D}[$\mathcal{A}^{acc}$]/$\mathcal{A}^{+}$[1]=$\varepsilon$::\texttt{*}]/$\mathcal{A}^{+}$[1]=$\varepsilon$::$A$].\hfill\(\Box\)\newline
\end{algorithmic}

Now, we generalize the formal definition of the algorithm \texttt{Rewrite} given in the previous section to handle predicates. Given the query \texttt{$p_{1}$/.../$p_{n}$} where 
\texttt{$p_{i}=axis_{i}$::$E_{i}$[$f_{i}$]} and $f_{i}$ ([$f_i$] is optional) is a predicate defined over element type $E_{i}$. We rewrite this query over node context of type \emph{E} as follows:

\begin{algorithmic}
\STATE \texttt{Rewrite$(p,E)$ := $\downarrow^{+}$::$E_{n}$[$\mathcal{A}^{acc}$][$f'_{n}$][$prefix^{-1}$($p_{1}$/.../$p_{n}$)]}
\end{algorithmic}

\noindent where an intermediate step, \texttt{$prefix^{-1}$($p_{1}$/.../$p_{i}$)} is recursively defined with:
\begin{itemize}
\item $axis_{i}$ = $\downarrow$: $prefix^{-1}$($p_{1}$/.../$p_{i}$) := $\mathcal{A}^{E_{i-1}}$[$f'_{i-1}$][$prefix^{-1}$($p_{1}$/.../$p_{i-1}$)]
\item $axis_{i}$ = $\downarrow^{+}$: $prefix^{-1}$($p_{1}$/.../$p_{i}$) := \newline$\uparrow^{+}$::$E_{i-1}$[$\mathcal{A}^{acc}$][$f'_{i-1}$][$prefix^{-1}$($p_{1}$/.../$p_{i-1}$)]\newline
where \texttt{$f'_{i-1}$ := RW\_Pred($f_{i-1},E_{i-1}$).}
\end{itemize}

\subsection{Complexity Analysis}
Given a security specification \emph{S}=(\emph{D},\texttt{ann}), we extract first the security view \emph{V}=$(D_{v},\texttt{ann})$ corresponding to \emph{S}, 
where $D_v$ is derived using our algorithm \texttt{DeriveView} of Figure \ref{Derive}. 
The user is allowed to request its authorized data represented with the DTD view $D_{v}$. 
For each query \emph{Q} in $\mathcal{X}$, our algorithm \texttt{Rewrite} translates this query to an equivalent one 
$Q_{t}$ in $\mathcal{X}^{\Uparrow}_{[n,=]}$ such that, for 
any instance \emph{T} conforms to \emph{D}, its virtual view $T_{v}$ conforms to $D_v$, we have $Q$($T_v$)=\texttt{Rewrite($Q$)}(\emph{T}).\newline

The overall complexity time of our rewriting algorithm \texttt{Rewrite} is stated as follows:

\begin{theorem}\label{OverallComplexity}
For any security view \emph{V}=$(D_{v},\emph{\texttt{ann}})$ and any $\mathcal{X}$ query \emph{Q} over the DTD view $D_{v}$, 
the algorithm \emph{\texttt{Rewrite}} computes an equivalent query $Q_{t}$ in $\mathcal{X}^{\Uparrow}_{[n,=]}$ over the original DTD in at most 
$O(|Q|*|D_{v}|^{2})$ time.\hfill\(\Box\)
\end{theorem}

\textsc{Proof.} Given an $\mathcal{X}$ query $p$=$axis_1$::$E_1$[$q_1$]/.../$axis_n$::$E_n$[$q_n$], we denote by $|p|$ the number of 
subqueries and sub-predicates of \emph{p}, e.g. |$\downarrow$::$E_1$[not($\downarrow^{+}$::\texttt{*})]|=2.
Each subquery (or sub-predicate) $p_i$=$axis_i$::$E_i[q_i]$ of \emph{p} must be rewritten over 
$p_{i-1}$=$axis_{i-1}$::$E_{i-1}[q_{i-1}]$ to check the accessibility of $E_i$ and to 
preserve the relationship defined between $E_i$ and $E_{i-1}$. This is done in a constant time by using the precomputed predicates $\mathcal{A}^{acc}$ and 
$\mathcal{A}^{+}$ as in lines 8-17 of algorithm \texttt{Rewrite} and lines 13-16 of algorithm \texttt{RW\_Pred}.
The \texttt{*}-label of each subquery (or sub-predicate) is eliminated as done in the lines 18-23 of algorithm \texttt{Rewrite}, and lines 3-8 of algorithm \texttt{RW\_Pred}, which 
causes an additional cost $|D_v|^{2}$. 
For instance, to rewrite the query $\downarrow$::\texttt{*} over the set \texttt{reach}, the elimination of the \texttt{*}-label amounts to parse each element type \emph{E} in \texttt{reach} and compute 
the union of its children types given by the precomputed list \texttt{Reach}($\downarrow$,$E$) (|\texttt{Reach}($\downarrow$,$E$)|=$O|D_v|$). 
Next, the \texttt{*}-label of the query $\downarrow$::\texttt{*} is replaced by the union of 
children types of element types in \texttt{reach} which is done in at most $O|D_v|^{2}$ time. 
Thus, a given query \emph{p} can be rewritten in at most $O(|p|*|D_{v}|^{2})$ time.\hfill\(\Box\)

\subsection{Query Rewriting Improvements}
We discuss in this section some possible implementations of our rewriting algorithm \texttt{Rewrite} to improve the overall complexity of Theorem \ref{OverallComplexity}.

The first optimization can be done by avoiding the \texttt{*}-label elimination step discussed above. 
For a security view \emph{V}=$(D_{v},\texttt{ann})$, an $\mathcal{X}$ query \emph{Q} over the DTD view $D_{v}$ can be rewritten in a linear time $O(|Q|)$.
Using the precomputed predicates $\mathcal{A}^{acc}$ and $\mathcal{A}^{+}$, 
each subquery $p_i$ of the query \emph{p} can be rewritten over $p_{i-1}$ in a \emph{constant time} by adding the 
predicate $\mathcal{A}^{E_{i-1}}$ (i.e. $\mathcal{A}^{+}$[1]/$\varepsilon$::$E_{i-1}$) or 
$\mathcal{A}^{acc}$, case of $axis_i$=$\downarrow$ and $axis_i$=$\downarrow^{+}$ respectively. For instance, 
the query $\downarrow$::\texttt{*}/$\downarrow$::$B$ over context node of type \emph{A} can be rewritten into 
$\downarrow^{+}$::$B$[$\mathcal{A}^{acc}$][$\mathcal{A}^{+}$[1]/$\varepsilon$::\texttt{*}/$\mathcal{A}^{+}$[1]/$\varepsilon$::$A$].
In the same way, each sub-predicate $q_i$=$axis_i$::$E_i$ can be rewritten over $q_{i-1}$ in a \emph{constant time} into 
$\downarrow^{+}$::$E_i$[$\mathcal{A}^{acc}$]/$\mathcal{A}^{+}$[1]=$\varepsilon$::$E_{i-1}$ in case of $axis_i$=$\downarrow$, or 
into $\downarrow^{+}$::$E_i$[$\mathcal{A}^{acc}$] otherwise.
For instance, the predicate [$\downarrow$::\texttt{*}/$\downarrow$::$C$] over element type \emph{B} 
can be rewritten into:
[$\downarrow^{+}$::\texttt{*}[$\mathcal{A}^{acc}$][$\downarrow^{+}$::$C$[$\mathcal{A}^{acc}$]/$\mathcal{A}^{+}$[1]=$\varepsilon$::\texttt{*}]/$\mathcal{A}^{+}$[1]=$\varepsilon$::$B$].
Thus, the rewriting of the query \emph{p} depends on the parsing of all its subqueries and sub-predicates which is done in $O(|p|)$ time.

Since the query answering time concerns the rewriting time and the evaluation time of the rewritten query, 
then the existence of the \texttt{*}-label in the rewritten query can induce for poor performance when 
some inaccessible elements are parsed by the rewritten query.
For instance, given the query $\downarrow$::$E$ over context node \emph{n} of type \emph{A} with $E$=\texttt{*}. Without eliminating the \texttt{*}-label, 
this query is rewritten into 
$\downarrow^{+}$::$E$[$\mathcal{A}^{acc}$][$\mathcal{A}^{A}$], and then for each descendant node of \emph{n} of any element type, the predicates 
$\mathcal{A}^{acc}$ and $\mathcal{A}^{A}$ are evaluated. To improve the query evaluation time, the elimination of the \texttt{*}-label in the query 
is indispensable and ensures that two defined predicates are evaluated only at accessible 
descendant nodes of \emph{n} (i.e. whose types appear in $D_v$).  This elimination can give good performance since the size of $D_v$ (number of accessible element types) is too 
small than the size of the original DTD \emph{D} in practice. 
For this reason, the precomputed lists \texttt{Reach} can be sorted in such a way the union of the children/descendant types of two element types of $D_v$ 
(e.g. \texttt{Reach}($\downarrow$,$E_1$) $\cup$ \texttt{Reach}($\downarrow$,$E_2$)) can be linear on the size 
of $D_v$. Accordingly, the \texttt{*}-label elimination phase, done in lines 18-23 of algorithm \texttt{Rewrite} and lines 3-8 of algorithm \texttt{RW\_Pred}, can be 
efficiently improved to take at most $|D_v|$ time and any query \emph{p} can be rewritten in this case in at most $O(|p|*|D_v|)$ time.

\section{Extensions}\label{extensions}
We discuss some extensions of our proposed rewriting approach to deal with a large fragment of XPath queries, and to 
overcome some limitations of existing access specifications languages.

\subsection{Upward-axes Rewriting}
For the rewriting of upward-axes ($\uparrow$ and $\uparrow^{+}$), we extend the algorithms \texttt{Rewrite} and \texttt{RW\_Pred} without increasing the complexity of the global 
rewriting (as explained in Theorem \ref{OverallComplexity}).\newline
In \texttt{Rewrite}, \texttt{$prefix^{-1}$($p_{1}$/.../$p_{i}$)} is defined over upward-axes as follows:

\begin{itemize}
\item \texttt{$axis_{i}=\uparrow$} : \texttt{$prefix^{-1}$($p_{1}$/.../$p_{i}$) := }\newline
\texttt{$\downarrow^{+}$::$E_{i-1}$[$\mathcal{A}^{acc}$][$f'_{i-1}$][$prefix^{-1}$($p_{1}$/.../$p_{i-1}$)]/$\mathcal{A}^{+}$[1]=$\varepsilon$::$E_{i}$}
\item \texttt{$axis_{i}=\uparrow^{+}$} : \texttt{$prefix^{-1}$($p_{1}$/.../$p_{i}$) := }\newline
\texttt{$\downarrow^{+}$::$E_{i-1}$[$\mathcal{A}^{acc}$][$f'_{i-1}$][$prefix^{-1}$($p_{1}$/.../$p_{i-1}$)]}\newline
where \texttt{$f'_{i}$ := \texttt{RW\_Pred}($f_{i},E_{i}$).}
\end{itemize}

\noindent In \texttt{RW\_Pred}, an intermediate predicate \texttt{$q_{i}$/../$q_{n}$}
(\texttt{$q_{i}$=$axis_{i}$::$E_{i}$[$f_{i}$]}) is rewritten over element type $E_{i-1}$ in case of upward-axes as follows:

\begin{itemize}
\item \texttt{$axis_{i}=\uparrow$}: \texttt{RW\_Pred($q_{i}$/.../$q_{n}$,$E_{i-1}$) := }\newline
\texttt{$\mathcal{A}^{E_{i}}$[$f'_{i}$][RW\_Pred($q_{i+1}$/.../$q_{n}$,$E_{i}$)]}
\item {$axis_{i}=\uparrow^{+}$}: \texttt{RW\_Pred($q_{i}$/.../$q_{n}$,$E_{i-1}$) := }\newline
\texttt{$\uparrow^{+}$::$E_{i}$[$\mathcal{A}^{acc}$][$f'_{i}$][RW\_Pred($q_{i+1}$/.../$q_{n}$,$E_{i}$)]}\newline
where \texttt{$f'_{i}$ := \texttt{RW\_Pred}($f_{i},E_{i}$).}
\end{itemize}

\subsection{Revision of Access Specifications}
\label{AccessSpecificationRevision}
The node accessibility w.r.t the specification value \texttt{[\emph{Q}]} has been defined with two different meanings \cite{ref4,ref6}.
Like in \cite{ref4}, we have assumed in the definition of the element node accessibility that for each element node 
\emph{n} concerned by an annotation of value [\emph{Q}], if \emph{Q} is not valid at \emph{n} (i.e. $n\nvDash Q$) then \emph{n} and all its descendant elements are not accessible, 
even if there is some valid annotations defined over these descendants (condition (\textit{ii}) of Definition \ref{accessibility}). 
However, in \cite{ref6} the element node \emph{n} can be not accessible (i.e. $n\nvDash Q$), but 
one of its descendant element $B'$ can be accessible if it is concerned by a valid annotation. 

We assume that both meanings are useful and an access control specification language must provide the definition of each of them.
For this reason, we redefine the access specification of Definition \ref{accessSpecification} as follows:

\begin{definition}\label{accessSpecifExtended}
An access specification \emph{S} is a pair $(D,\emph{\texttt{ann}})$ consisting of a DTD \emph{D} and a partial mapping \emph{\texttt{ann}} such that, for each production $A \rightarrow P(A)$ and each element type \emph{B} in \emph{P(A)}, 
\emph{\texttt{ann}(A,B)}, if explicitly defined, is an annotation of the form: 
\begin{center}
 \emph{\texttt{ann($A$,$B$) :=  $Y$ | $N$ | [$Q$] | $N_h$ | [$Q$]$_h$}}
\end{center}
\hfill\(\Box\)
\end{definition}

\noindent Given an element node \emph{n} of type \emph{B} with parent node of type \emph{A}, 
then with the specification values \emph{N} and [\emph{Q}], accessibility overwriting is allowed under
\emph{n} even though \texttt{ann}(\emph{A,B})=\emph{N} or \texttt{ann}(\emph{A,B})=[\emph{Q}] and $n\nvDash Q$.
The semantics of the new specification values $N_{h}$ and \texttt{[\emph{Q}]$_{h}$} are given as follows. 
If the element node \emph{n} is concerned by an annotation with value $N_h$, then no overwriting of this value is permitted to descendant elements of \emph{n}, 
i.e. if $n'$ is a descendant element of \emph{n}, then $n'$ is not accessible even if it is concerned by a valid annotation.
While if \emph{n} is concerned by an annotation with value [$Q$]$_{h}$, then the annotations defined under \emph{n} (i.e. under \emph{B} element type) 
take effect only if $n\vDash Q$. For instance, if a descendant element $n'$ of \emph{n} is concerned by an annotation of value $[Q']$, then $n'$ is accessible 
only if $n'\vDash Q'$ and $n\vDash Q$. We call the annotation with value $N_{h}$ or [Q]$_{h}$, \emph{downward-closed} annotation.\newline

\begin{figure}[t!]
\centering
\includegraphics[width=8cm]{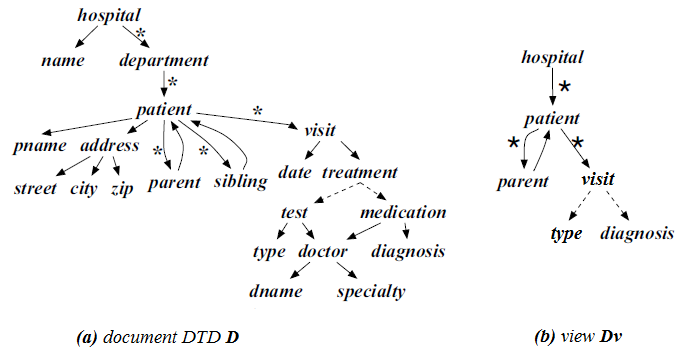}
\caption{Hospital DTD.}
\label{hospitalDTD}
\end{figure}

\noindent \textbf{Example 5.1} We consider the hospital DTD of Figure \ref{hospitalDTD}(a) and we give the 
following access specification:

\begin{itemize}
\item \textit{\textbf{A patient can access only to its own $diagnosis$ information:}}\newline
\texttt{ann$(department,patient)$ = [$pname$=$\$name$]$_{h}$}\newline
\texttt{ann$(patient,parent)$ = ann$(patient,sibling)$=$N_{h}$}\newline
\texttt{ann$(patient,visit)$ = $N$, ann$(medication,diagnosis)$ = $Y$}
\item \textit{\textbf{A research institute can access to patients whose have "$disease1$":}}\newline
\texttt{ann$(department,patient)$ = ann$(parent,patient)$ = }\texttt{ann$(sibling,patient)$}\newline
\texttt{= [$visit$/$treatment$/$medication$[$diagnosis$='disease1']]}
\end{itemize}

\noindent For the first policy, $\$name$ is a variable system denoting patient's name. For a given patient with name $"Bob"$, if $p\nvDash pname$=$"Bob"$
then all the fragment rooted at $p$ is hidden from 
the patient $"Bob"$ and any annotation under element node $p$ can be applied since it is the medical data of another patient. 
Also, the annotation $\texttt{ann}(patient,parent)$=$N_h$ cannot be defined with $\texttt{ann}(patient,parent)$=$N$ 
otherwise $\texttt{ann}(medication,diagnosis)$=$Y$ makes diagnosis information of parent accessible to patient $"Bob"$ as its own information. 
The same principle is applied with the annotation $\texttt{ann}(patient,sibling)$=$N_h$.

\noindent For the second policy, it is clear that a false evaluation of the predicate does not imply the inaccessibility of the sub patients, i.e. a 
given patient may not have "\emph{disease1}", while 
its parent/sibling can be affected by this disease and must be shown to the research institute.\hfill\(\Box\)\newline

The new access specification defined can be taken into account simply by applying the following changes in our rewriting approach. 
The predicate $\mathcal{A}_1^{acc}$ is redefined with:\footnote{Note that $A.\sigma(A',A)$ gives $A[Q]$ if \texttt{ann($A'$,$A$)=[$Q$]|[$Q$]$_h$} and \emph{A} otherwise.}

\begin{algorithmic}
\STATE \texttt{$\mathcal{A}_1^{acc}$ := $\uparrow^{*}$::*[$\varepsilon$::$root$ $\vee_{ann(A',A)\in ann}$ $\varepsilon$::$A$/$\uparrow$::$A'$][1]}
\STATE \texttt{[$\varepsilon$::$root$ $\vee_{(ann(A',A)=Y|[Q]|[Q]_h)\in ann}$ $\varepsilon$::$A.\sigma(A',A)$/$\uparrow$::$A'$]}
\end{algorithmic}

\noindent While The predicate $\mathcal{A}^{acc}_2$ is redefined with:

\begin{algorithmic}
\STATE \texttt{$\mathcal{A}_2^{acc}$:= $\bigwedge_{(ann(A',A)=N_h)\in ann}$ not ($\uparrow^{+}$::$A$/$\uparrow$::$A'$)}
\STATE \texttt{$\bigwedge_{(ann(A',A)=[Q]_h)\in ann}$ not ($\uparrow^{+}$::$A$[not (\emph{Q})]/$\uparrow$::$A'$)}
\end{algorithmic}

\section{Experimental Results}\label{Sect6}
We have developed a prototype to improve effectiveness of our rewriting approach. The performance study is done using a real-life recursive DTD 
and a various forms of XPath queries.
The experimental results show the efficiency of our XPath query rewriting approach w.r.t the answering approach 
based on the materialization of the view. Notice that we cannot do comparison between our approach and the 
two existing approaches which deal with queries rewriting under recursive views \cite{ref5,ref8}, since they are based on 
the non-standard language "regular XPath", and no practical tool is present to evaluate regular XPath queries. 
The experiments were conducted using Ubuntu system, with a dual Core 2.53 GHz and 1 GB of memory.\newline

\noindent \textbf{XML Documents.} 
Using \texttt{ToXGene} generator \cite{ref9}, we generated set of XML 
documents that conform to the hospital DTD of Figure \ref{hospitalDTD} and with sizes ranging from 10MB to 100MB.

\noindent \textbf{Security Specification.} Figure \ref{hospitalDTD}(b) represents the hospital DTD view $D_{v}$ of a research institute studying inherited patterns of some 
diseases. This view shows only 
patients having one or more disease from \{\emph{disease1, disease2, disease3}\} with their parent hierarchy, and denies access to their 
\emph{name}, \emph{address}, \emph{test} and \emph{doctor} data. Formally, we define this view with the following annotations:

\begin{itemize}
\item[1.] \texttt{ann($hospital$)=$Y$}
\item[2.] \texttt{ann($hospital$,$name$)=$N$}
\item[3.] \texttt{ann($hospital$,$department$)=$N$}
\item[4.] \texttt{ann($department$,$patient$)=}\newline \texttt{[$\downarrow$::$visit$/$\downarrow$::$treatment$/$\downarrow$::$medication$[$\downarrow$::$diagnosis$='disease1' or \newline
				    $\downarrow$::$diagnosis$='disease2' or $\downarrow$::$diagnosis$='disease3']]$_{h}$}
\item[5.] \texttt{ann($patient$,$pname$)=$N$}
\item[6.] \texttt{ann($patient$,$address$)=$N$}
\item[7.] \texttt{ann($patient$,$sibling$)=$N_{h}$}
\item[8.] \texttt{ann($visit$,$date$)=$N$}
\item[9.] \texttt{ann($visit$,$treatment$)=$N$}
\item[10.] \texttt{ann($medication$,$diagnosis$)=$Y$}
\item[11.] \texttt{ann($test$,$type$)=$Y$}
\end{itemize}

\noindent The annotation 7 must be downward-closed, otherwise the annotations 10 and 11 can overwrite some sibling data (\emph{diagnosis} and \emph{type} of visit) to be 
accessible.\newline

\noindent \textbf{XPath Queries.} We define the following set of XPath queries :

\begin{figure}[H]
\begin{enumerate}
 \item \texttt{$Q_{1}$. $\downarrow$::$patient$[$\downarrow^{+}$::$visit$[$\downarrow$::$diagnosis$='disease1' or \newline
				    $\downarrow$::$diagnosis$='disease2' or $\downarrow$::$diagnosis$='disease3']]}.
 \item \texttt{$Q_{2}$. $\downarrow^{+}$::$patient$[$\downarrow$::$visit$[$\downarrow$::$diagnosis$='disease1' or $\downarrow$::$diagnosis$='disease2' or $\downarrow$::$diagnosis$='disease3'] and \newline
                not ($\downarrow^{+}$::$patient$/$\downarrow$::$visit$[$\downarrow$::$diagnosis$='disease1' or $\downarrow$::$diagnosis$='disease2' or $\downarrow$::$diagnosis$='disease3'])]}.
 \item \texttt{$Q_{3}$. $\downarrow^{+}$::$diagnosis$[$\uparrow$::$visit$/$\uparrow$::*/$\uparrow$::*/$\uparrow$::*/$\uparrow$::$hospital$]}
\end{enumerate}
\normalsize
\label{dtdView}
\end{figure}

\noindent The first query returns patients whose some of its ancestors also had the sames diseases. The Second query $Q_{2}$ returns the first generation 
where the discussed diseases appeared for the first time, and $Q_{3}$ represents the diagnosis of the second generation of infected patients. 
Each query $Q_i$ is rewritten over the root node (\emph{hospital}) of each document into \texttt{Rewrite}($Q_i$,$hospital$), and this by using the security view \emph{V}=$(D_v,\texttt{ann})$ 
defined with the DTD view of Figure \ref{hospitalDTD}(b) and the annotations defined above.

\noindent \textbf{Approaches.} A comparison is done between our rewriting approach and the \emph{materialization} approach.
Given a security view \emph{V}=$(D_v,\texttt{ann})$, the \emph{materialization} consists in incorporating 
an accessibility label (+/-) to each element node in the document which is concerned by an annotation of \emph{V}. 
Each element node \emph{n}, not yet labeled (i.e. no invalid downward-closed annotation is defined over its ancestor elements), 
is labeled with ``+'' if it is concerned by an annotation with value \emph{Y} or with value $[Q]|[Q]_h$ and $n\vDash Q$ 
(resp. \emph{n} is labeled with ``-'' in case of annotation with 
value $N$|$N_h$ or with value $[Q]|[Q]_h$ and $n\nvDash Q$).
In case of an element node \emph{n} concerned by an invalid downward-closed annotation (with value $N_h$ or with value $[Q]_h$ and $n\nvDash Q$), the 
\emph{n} and all its descendant elements are labeled with ``-''. 
After applying all the annotations of the security view \emph{V} over the document, each unlabeled element node is annotated by inheritance 
from its nearest labeled ancestor element.
The obtained document is called \emph{fully annotated document}. Finally, the \emph{materialized view} of the 
original document is computed by deleting all inaccessible element nodes (labeled with ``-'') from the fully annotated document and user 
queries are evaluated directly over this view. Thus, 
we compare the answering time of the materialization approach (defined 
as the \emph{view materialization} time and \emph{query evaluation} time over the materialized view) with that of our rewriting approach (defined 
as the rewriting time of the query and the evaluation time of the rewritten query over the original document).\newline

\begin{figure*}[t!]
\centering
\includegraphics[width=12cm, height=4.4cm]{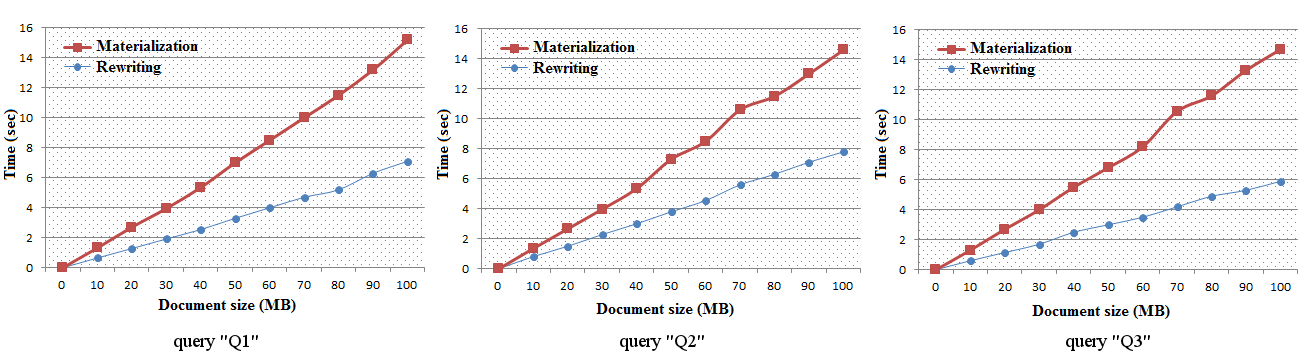}
\caption{XPath Queries Evaluation Time.}
\label{performanceStudy}
\end{figure*}

\noindent \textbf{Performance Results.}
The experimental results are shown in Figure \ref{performanceStudy} where the answering time of each query is evaluated using our rewriting algorithm and the 
materialization approach. The size of the answer ranges from few hundred to a few thousand of nodes. 
Figure \ref{performanceStudy} shows clearly that our algorithm remains more efficient than the materialization approach.

We observe first that the translation of XPath queries from $\mathcal{X}$ to $\mathcal{X}^{\Uparrow}_{[n,=]}$ does not induce 
for a poor performance and the average of the answering time of our rewriting approach remains in general less than 8 seconds for a large XML document. 
Second, a query containing time-consuming elements like \texttt{*}-labels or \texttt{parent} axes does not degrade the rewriting performance as shown with 
the query $Q_3$.

\section{Conclusions and Future Work}\label{Sect7}
The proposed approach yields the first practical solution to rewrite XPath queries over recursive XML views using only the expressive power of the standard XPath. 
The extension of the downward class of XPath queries with some axes and operators has been investigated in order to make queries rewriting possible under 
recursion. 

The conducted experimentation shows the efficiency of our approach by comparison with the 
materialization approach. Most importantly, the translation of queries from $\mathcal{X}$ to $\mathcal{X}^{\Uparrow}_{[n,=]}$ does not impact the performance 
of the queries answering.
We have discussed how our approach can be extended to deal with the upward-axes without additional cost. Lastly, 
a revision of the access specification language is presented to go beyond some limitations in the definition of some access privileges.

As future work, we plan first to provide an optimized version of our approach and also to use the same principle to secure XML updating.

\bibliographystyle{abbrv} 
\bibliography{RR-7834}

\begin{thebibliography}{10}

\bibitem{ref9}
D.~Barbosa, A.~Mendelzon, J.~Keenleyside, and K.~Lyons.
\newblock Toxgene: An extensible template-based data generator for xml.
\newblock {\em In WebDB 2002, pp. 49-54}.

\bibitem{XPath20}
A.~Berglund, S.~Boag, D.~Chamberlin, M.~F. Fern\'andez, M.~Kay, J.~Robie, and
  J.~Sim\'eon.
\newblock Xml path language (xpath) 2.0 (second edition). w3c recommendation 14
  december 2010. http://www.w3.org/tr/2010/rec-xpath20-20101214/.

\bibitem{ref22}
E.~Bertino and E.~Ferrari.
\newblock Secure and selective dissemination of xml documents.
\newblock {\em In TISSEC}, 5(3):290--331, 2002.

\bibitem{ref4}
W.~Fan, C.-Y. Chan, and M.~Garofalakis.
\newblock Secure xml querying with security views.
\newblock {\em In SIGMOD 2004, pp. 587-598.}

\bibitem{ref5}
W.~Fan, F.~Geerts, X.~Jia, and A.~Kementsietsidis.
\newblock Rewriting regular xpath queries on xml views.
\newblock {\em In ICDE 2007, pp.666-675.}

\bibitem{ref15}
W.~Fan, F.~Geerts, X.~Jia, and A.~Kementsietsidis.
\newblock Smoqe: A system for providing secure access to xml.
\newblock {\em In VLDB 2006, pp. 1227-1230.}

\bibitem{ref2}
I.~Fundulaki and M.~Marx.
\newblock Specifying access control policies for xml documents with xpath.
\newblock {\em In SACMAT 2004, pp. 61-69.}

\bibitem{ref8}
B.~Groz, S.~Staworko, A.-C. Caron, Y.~Roos, and S.~Tison.
\newblock Xml security views revisited.
\newblock {\em In DBPL 2009, pp, 52-67}.

\bibitem{ref6}
G.~Kuper, F.~Massacci, and N.~Rassadko.
\newblock Generalized xml security views.
\newblock {\em In SACMAT 2005, pp. 77-84.}

\bibitem{ref1}
M.~Murata, A.~Tozawa, and M.~Kudo.
\newblock Xml access control using static analysis.
\newblock {\em In CCS 2003, pp. 73-84.}

\bibitem{ref7}
N.~Rassadko.
\newblock Policy classes and query rewriting algorithm for xml security views.
\newblock {\em In Data and Applications Security 2006, pp. 104-118.}

\bibitem{ref13}
N.~Rassadko.
\newblock Query rewriting algorithm evaluation for xml security views.
\newblock {\em In Secure Data Management 2007, pp. 64-80.}

\bibitem{ref14}
A.~Stoica and C.~Farkas.
\newblock Secure xml views.
\newblock {\em In the IFIP WG 11.3 Sixteenth International Conference on Data
  and Applications Security 2002, pp. 133-146.}

\bibitem{ExpressivityXPath}
B.~ten Cate.
\newblock The expressivity of xpath with transitive closure.
\newblock {\em In PODS 2006, pp. 328-337}.

\bibitem{ComplexityXPath20}
B.~ten Cate and C.~Lutz.
\newblock The complexity of query containment in expressive fragments of xpath
  2.0.
\newblock {\em Journal of the ACM 2009, pp. 31-48}.

\bibitem{ref12}
R.~Vercammen, J.~Hidders, and J.~Paredaens.
\newblock Query translation for xpath-based security views.
\newblock {\em In EDBT 2006, pp. 250-263.}

\end{thebibliography}
\nocite{ref14,ref22}
\end{document}